\documentclass[acmlarge,authorversion]{acmart}

\AtBeginDocument{%
  \providecommand\BibTeX{{%
    \normalfont B\kern-0.5em{\scshape i\kern-0.25em b}\kern-0.8em\TeX}}}

\usepackage{booktabs,caption}
\usepackage[flushleft]{threeparttable}
\usepackage{soul}
\usepackage{diagbox}
\usepackage{multirow}

\setcopyright{acmlicensed}
\acmJournal{CSUR}
\acmYear{2023} \acmVolume{1} \acmNumber{1} \acmArticle{1} \acmMonth{1} \acmPrice{15.00}\acmDOI{10.1145/3606947}

\begin{document}

\title{An End-to-End Review of Gaze Estimation and its Interactive Applications on Handheld Mobile Devices}

\author{Yaxiong Lei}
\email{yl212@st-andrews.ac.uk}
\orcid{0000-0002-0697-7942}
\affiliation{%
  \institution{University of St Andrews}
  \city{St Andrews}
  \country{UK}
  \postcode{KY16 9SX}
}

\author{Shijing He}
\email{shijing.he@kcl.ac.uk}
\orcid{0000-0003-3697-0706}
\affiliation{%
 \institution{King's College London}
 \city{London}
 \country{UK}}
 
\author{Mohamed Khamis}
\email{mohamed.khamis@glasgow.ac.uk}
\orcid{0000-0001-7051-5200}
\affiliation{%
 \institution{University of Glasgow}
 \city{Glasgow}
 \country{UK}}
 
\author{Juan Ye}
\email{jy31@st-andrews.ac.uk}
\orcid{0000-0002-2838-6836}
\affiliation{%
 \institution{University of St Andrews}
  \city{St Andrews}
  \state{Fife}
  \country{UK}
  \postcode{KY16 9SX}
}

\renewcommand{\shortauthors}{Y. Lei et al.}

\begin{abstract}
In recent years we have witnessed an increasing number of interactive systems on handheld mobile devices which utilise gaze as a single or complementary interaction modality. This trend is driven by the enhanced computational power of these devices, higher resolution and capacity of their cameras, and improved gaze estimation accuracy obtained from advanced machine learning techniques, especially in deep learning. As the literature is fast progressing, there is a pressing need to review the state of the art, delineate the boundary, and identify the key research challenges and opportunities in gaze estimation and interaction. This paper aims to serve this purpose by presenting an end-to-end holistic view in this area, from gaze capturing sensors, to gaze estimation workflows, to deep learning techniques, and to gaze interactive applications. 
\end{abstract}





\maketitle

% ======== Main Content ========%

\section{Introduction}
Gaze interaction is to make use of gaze to facilitate interactions with computing devices, including virtual reality (VR) headsets, mobile or wearable devices, desktops, and robots. Gaze refers to a point on a screen or a direction in space and can be inferred from pupil positions, facial structure, and head movements. Through gaze, a system can sense users' attention~\cite{van2018gazepath} and enable touch-free interaction, which has driven a wide range of applications. 

In a simulated surgical training task, gaze has been used to aid the identification of a target on a subject's laparoscopic screen in order to reduce mistakes and overcome language barriers~\cite{chetwood2012collaborative}. In a driving situation, real-time gaze coding allows to track a driver's attention and detecting whether they are distracted or fatigue~\cite{jha2022estimatedrivergaze,yang2020all}, and support automated driving~\cite{mole2021drivers}. Gaze can also help diagnose mental health or autism by analysing the scan path~\cite{venuprasad2019characterizing}. Moving beyond, the gaze is also presented as a human-computer interface, facilitating controlling Internet of Things (IoT) devices in a smart home system~\cite{kim2019watch}, and empowering people with a physical impairment to interact with applications such as creative art~\cite{zhang2019eye, eid2016eyechair, krishna2020eyerobotic}. 

The advanced research in gaze interaction has inspired an increasing number of industrial applications, including optimising the system's scheduling of processing resources~\cite{mathur2021resource},  facilitating the presentation of content~\cite{kim2016understanding, feit2020detecting}, marketing, and accessibility for people with disabilities~\cite{creed2020multimodal, hemmingsson2020usability, zhang2017smartphone, fan2020eyelid}. Smart Eye brings an AI-integrated eye-tracking technology to detect whether a driver is  distracted~\cite{smarteye2023}, and Eyetech Digital Systems launched EyeOn to enable users to type and communicate using eye movement~\cite{eyetech2023}. There are growing market players, including Tobii AB, LC Technologies, Eyetech Digital Systems, Ergoneers GmbH, Smart Eye AB., Mirametrix Inc., Pupil Labs GmbH, Seeing Machines, SR Research Ltd., and Gazepoint. The worldwide market for eye tracking technology is valued at USD 638.8 million and is anticipated to grow at a compound annual growth rate (CAGR) of 33.4\% between 2022 and 2030~\cite{grandviewresearch2022}.

In recent years, mobile devices have advanced with significantly improved camera quality and computational power. These capabilities make it possible to explore gaze interaction on mobile devices. The advantage over traditional eye tracking is that gaze on mobile devices is derived from images or videos on cameras, which does not require extra devices. Industries are starting to explore novel gaze interactions on mobile devices. For example, Apple has paired facial recognition and eye movements to enhance the unlocking experience of FaceID and the Huawei Mate 40 Pro keeps the screen on when being gazed at.

What currently needs to be added to the literature is a comprehensive overview of gaze estimation to interaction on handheld mobile devices and a roadmap for advancing from low-level gaze points and directions to high-level gaze patterns, which play a significant role in many gaze applications. To close the gap, we describe the current technological advances in capturing gaze and a workflow along with deep learning algorithms to estimate gaze. We identify the challenges in enhancing the diversity of data collection and robustness of gaze estimation,  especially reflecting the complexities of dynamic environments such as partial faces, varying lighting conditions, and constant changes of holding postures. These challenges come from the inherent characteristics of mobile devices: their size and mobility, which fundamentally differ from eye tracking on other platforms such as VR, desktops, and large displays. We also broadly review gaze-based applications, introduce eye physiology, and point to new types of gaze for future interaction design. Different from the existing surveys on gaze estimation, this paper makes the following key contributions.
\begin{itemize}
    \item We present an end-to-end holistic review from sensors, to algorithms, and to application.
    \item We focus on appearance-based gaze estimation on handheld devices, and identify the unique challenges different from other platforms.
    \item We bridge the gap between gaze estimation, gaze data processes, and gaze interaction and present a pipeline for processing, analysing, and deriving high-level gaze events from raw gaze data. 
\end{itemize}

\section{Methodology}\label{sec:method}
Our paper takes recent reviews and survey studies as an initial research methodology~\cite{khamis2018past,katsini2020role,cheng2021appearance,hirzle2020survey,kar2017review}, with adaptation to suit our research aims; that is, understanding the workflow of gaze interaction from camera input to applications on handheld mobile devices. Our research methodology comprises four stages: defining keywords, paper collection, paper categorisation and collation, and paper analysis. We extract keywords based on the content in the papers mentioned in the related work, and eventually identify (``Eye'' or ``Gaze'' or ``Eye Tracking'') and (``Gaze Estimation'' or ``Eye Tracking Technique'' or ``Eye Tracking algorithm'' or ``Algorithm" or ``Dataset'') for Section~\ref{sec:overview} and \ref{sec:datasets}, (``Eye" or ``Gaze'' or ``Eye Tracking'' or ``Eye Movement'') and (``Data Analysis'' or ``Data Process'' or ``Application'' or `` Interaction'') for Section~\ref{sec:gazeanalysis} and \ref{sec:application}. We select the conferences and journals in the areas of HCI, UbiComp, computer vision, and eye-tracking as our target sources, including ETRA, CHI, IJHCS, HCI, UbiComp/IMWUT, MobileHCI, IJHCI, CVPR, EMR, BRM, etc. We also search terms on Google Scholar to catch more publications.

We categorise papers based on research topics: gaze estimation algorithms, gaze data analysis, gaze interaction and gaze applications. We then sub-categorise them according to the platforms in which these items are applied, e.g., handheld mobile devices (mobile phones or tablets), desktop devices (desktop computers or laptops), and head-mounted devices (VR or glasses). Algorithms, interactions and applications that are currently used in various devices will be discussed in terms of their applicability to handheld mobile devices. Our discussion focuses on the characteristics of handheld mobile devices, i.e. the mobility of the device, the screen size, the application characteristics, and the usage scenario.

Based on the classification and year, we summarise the content of the collected papers, extract their motivation, methodology, subsequent research directions and their contribution to gaze estimation algorithms and applications for handheld mobile devices.

\section{Related Work}\label{sec:related}
In recent years, more and more research projects on gaze are emerging, addressing the quest for performance and stability in gaze estimation, gaze interaction and gaze applications using the capabilities of commercial eye trackers or gaze estimation algorithms. The existing surveys and review studies have different perspectives and areas: algorithms and applications of gaze estimation. Kar and Corcoran~\cite{kar2017review} survey nearly 20 years of gaze estimation research on a variety of devices, including desktop, television, head-mounted, automotive and handheld, and also focus on methods for evaluating the performance of gaze tracking systems. Cheng et al.~\cite{cheng2021appearance} provide a detailed review of appearance-based gaze estimation approaches. They build a pipeline of gaze estimation of deep learning. In terms of applications, Khamis et al.~\cite{khamis2018past} summarise eye-tracking related studies on handheld mobile devices from 2002 to 2018. They list three main topics of gaze applications: gaze behaviour analysis, implicit gaze interaction and explicit gaze interaction. Katsini et al.~\cite{katsini2020role} focus on the application of gaze in security and privacy, which include authentication, privacy protection and gaze monitoring during security critical tasks. They summarise the usage scenarios, devices and evaluation methods for these tasks. Hirzle et al.~\cite{hirzle2020survey} conducted a study on existing gaze interaction projects around the Digital Eye Strain (DES), which has surveyed over 400 papers published in the last 46 years. Ghosh et al.~\cite{ghosh2022automatic} focus on the algorithms in AR/VR devices and the applications in healthcare and driver engagement. Nishan et al.~\cite{nishan2022eyeedge} provides a detailed analysis of eye-tracking technology on mobile devices and present an edge computing solution for real-time eye-tracking experience. 

Different from the above reviews, our paper aims to present an end-to-end overview of gaze estimation, data analysis and interaction, illustrating its workflow, mainstream deep learning techniques for estimating gaze, and a broad range of gaze interaction applications on handheld mobile devices.

\section{Gaze Estimation on Handheld Mobile Devices}\label{sec:overview}
This section presents an overview of gaze estimation process, including the task of gaze estimation (in Section~\ref{subsec:overview}), sensors common in handheld mobile devices for gaze estimation (in Section~\ref{subsec:sensors}), and the workflow (in Section~\ref{subsec:workflow}) including \textit{pre-processing}, \textit{learning}, \textit{post-processing}, and \textit{calibration}. 

\subsection{Task and Evaluation of Gaze Estimation}\label{subsec:overview}
\begin{figure}[!htbp]
    \centering
    \includegraphics[width=0.5\textwidth]{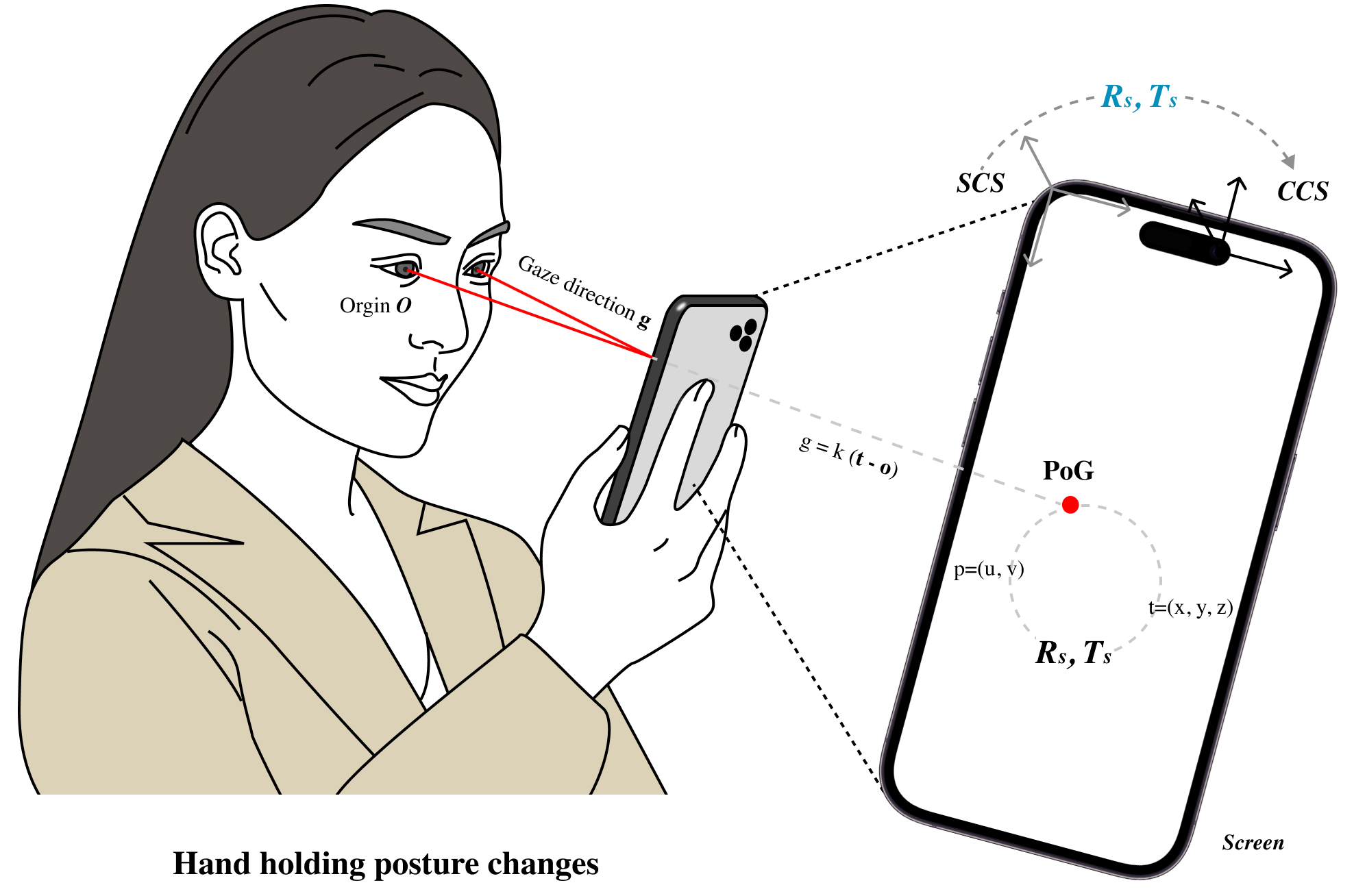}
    \caption{The gaze estimation task involves predicting either a) a point of gaze (PoG) as a coordinate point on a screen (called 2D gaze) or b) a gaze direction in a 3D space (called 3D gaze)~\cite{cheng2021appearance}. }
    \label{fig:estimation}
\end{figure}

Gaze estimation refers to predicting a point of gaze (PoG) \cite{cheng2021appearance} from images or videos captured on the front camera of handheld mobile devices, as presented in Figure~\ref{fig:estimation}; i.e., a coordinate point $p=(u,v)$  on a screen; or gaze direction; i.e., a vector $\mathbf{g}=(g_x, g_y, g_z)$ in a 3D coordinate system \cite{mora2013person}. The former is often referred to as 2D gaze while the latter as 3D gaze. In both cases, it is treated as a regression task, while it can also be treated as a classification problem~\cite{vora2017generalizing}; that is, predicting which grid or area the user is looking at. 

The accuracy of gaze estimation is defined as the average difference between the estimated gaze location and the location of the fixation target. In 2D, the difference is measured as the Euclidean distance between the true point $\boldsymbol{p}$ and estimated point $\hat{\boldsymbol{p}}$, where a point is commonly represented as a 2D coordinate in pixels with units of cm or mm~\cite{krafka2016eye, bao2021adaptive, chang2021high, park2021gazel}.
\begin{equation}\label{eq:2D}
    acc_{2D}= d = {\|\boldsymbol{p} - \hat{\boldsymbol{p}}\|}
\end{equation}

The true points can be obtained via an extra high-resolution eye-tracking device or self-reported by participants; that is, indicate their gaze points using a cursor or touch~\cite{liebling2014gaze, kasprowski2016implicit, huang2016autocalibration, jiang2020howtype, weill2016you}. For example, GazeCapture~\cite{krafka2016eye} designs an application to display a red dot on a screen of a mobile phone and requires participants to gaze at the dot and follow its movement. Before the dot moves, a small letter L or R is displayed for 0.05 seconds, which requires a participant to tap either the left (L) or right (R) side of the screen. This serves as a way to monitor participants' attention and thus validate the data collection. The ground truth points are the locations of the red dot. 

In 3D, the difference in gaze direction refers to an angular distance between the true direction vector $\boldsymbol{g }$ and the estimated direction $\hat{\boldsymbol{g }}$, which is often represented in degrees. 
\begin{equation}\label{eq:3D}
    acc_{3D}= \alpha^{\circ}=\frac{\boldsymbol{g } \cdot \hat{\boldsymbol{g }}}{\|\boldsymbol{g }\|\|\hat{\boldsymbol{g}}\|}
\end{equation}

The collection of gaze direction is more complex, as the ground truth is a vector in a 3D space. Gaze360~\cite{kellnhofer2019gaze360} uses a 360$^{\circ}$ panoramic camera placed on a tripod in the centre of the scene, and a large moving rigid target board marked with an AprilTag and a cross on which participants are instructed to continuously fixate. The true gaze direction is derived from the camera coordinator system based on the distance between the eye, the camera, and the target board. Differently, ETH-XGaze~\cite{zhang2020eth} utilises 18 single-lens reflex cameras in a 3D space to gather ground truth while stimuli are displayed on a large screen in front of participants.

Commercial eye-tracking technologies like Tobii~\cite{tobii2020test} also provide the other two measures: precision and data loss. Precision refers to the system's ability to produce the same gaze point or direction measurement reliably. It is calculated as the root mean square of accuracy on a sequence of consecutive data pairs of true and estimated gazes. 

\begin{equation}
    precision = \sqrt{\frac{\sum^n_{i=1} acc^2_i}{n}}, 
\end{equation}
where $acc_i$ is the accuracy (2D or 3D) on the $i$th pair between the target location and the gaze (point or direction), referring to the above Equation~(\ref{eq:2D}) and (\ref{eq:3D}).

Data loss is defined as the ratio of gaze samples captured by the eye tracker during the fixation on a target. It is calculated by excluding invalid samples; that is, no gaze is detected. 
\begin{equation}
    data\_loss = \frac{\text{No. of invalid samples}}{\text{No. of total samples}}
\end{equation}

Current gaze estimation algorithms typically focus on providing accuracy metrics and do not include precision and data loss measurements. However, for practical applications, it is important to consider all the above three metrics and provide them in their test report to ensure reliable and stable performance in real-world applications. 

\subsection{Sensors}\label{subsec:sensors}
\begin{figure}[!htbp]
    \centering
    \includegraphics[width=\columnwidth]{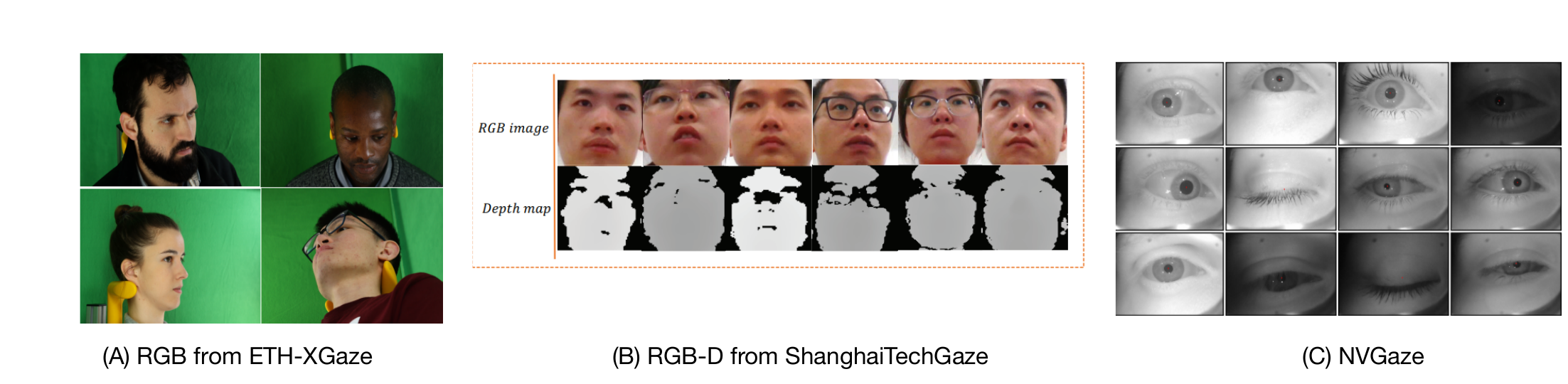}
    \caption{Examples of images from cameras that are readily integrated into handheld mobile devices, and can be used for gaze estimation: (A) RGB~\cite{zhang2020eth}, (B) RGB-D~\cite{lian2019rgbd}, and (C) IR~\cite{kim2019nvgaze} }
    \label{fig:sensors}
\end{figure}

Eye-tracking technologies have advanced significantly over the past 60 years~\cite{shackel1960pilot}, from electro-oculoGraphy (EOG) signals that are based on muscle action signals to visual signals from cameras. The recent examples include head-mounted devices~\cite{beach1998eye} like Oculus VR or HoloLens 2, wearable eye trackers like Tobii Pro Glasses 3, and various types of screen-based eye trackers~\cite{park2020towards}. These devices come in a variety of shapes but generally tightly integrate the camera above or below a screen; for example, a camera at the top of a mobile phone. They have different eye-to-screen distances; for example, head-mounted eye trackers are close to the eye (e.g., within the distance of 0.5-3cm), while in the screen-based eye trackers (aka remote eye trackers), the distance is usually 50-80cm for desktop settings~\cite{tobiipro_2015}, and 15-50cm for handheld devices~\cite{huang2017screenglint,valliappan2020accelerating}. The distance and the screen size of handheld devices may constrain what types of gaze interaction applications are more acceptable to users. 
In this survey, we focus on sensors on handheld mobile devices, which are RGB, RGB-D(epth), and Infrared (IR) cameras (see Figure~\ref{fig:sensors}).

\subsubsection{RGB Camera}
Currently, most mobile device-based gaze estimation methods are using RGB Cameras, which are widely supported. Over years, the quality of the cameras has improved, in terms of aperture, resolution, and frame rate.

\subsubsection{RGB-D Camera}
RGB-D camera adds depth information to RGB images; that is, each pixel relates to a distance between the object in the image and the image plane. RGB-D cameras are starting to be supported in modern smartphones, especially high-end phones. Depth information can assist in constructing a 3D model of head pose and provide information on the eye position, further improving gaze estimation~\cite{mora2013person, xiong2019mixed}. 

\subsubsection{IR Camera}
IR camera refers to a Near Infrared (NIR) camera~\cite{prokoski2000history} that uses an artificial IR light source aimed on- and off-axis at the eye, which introduces \textit{glint}, called corneal reflection~\cite{majaranta2014eye}. The glint acts as a reference point and the gaze direction is calculated by measuring the changing relationship between the glint and the moving pupil centre of the eye. IR camera is sensitive to wavelengths from 700 nanometers (nm) to 1,400 nm, and this band does not affect human vision beyond the range of visible spectrum~\cite{starr2014biology}, so it is often used as near-eye cameras; e.g., head-mounted augmented and virtual reality (AR/VR) devices~\cite{wu2019eyenet}. Recently, there are more and more mobile phones that are configured with NIR cameras, including iPhone, Huawei, Samsung, Xiaomi and OPPO. 

\subsection{Workflow}\label{subsec:workflow}
\begin{figure}[!htbp]
    \centering
    \includegraphics[width=0.9\textwidth]{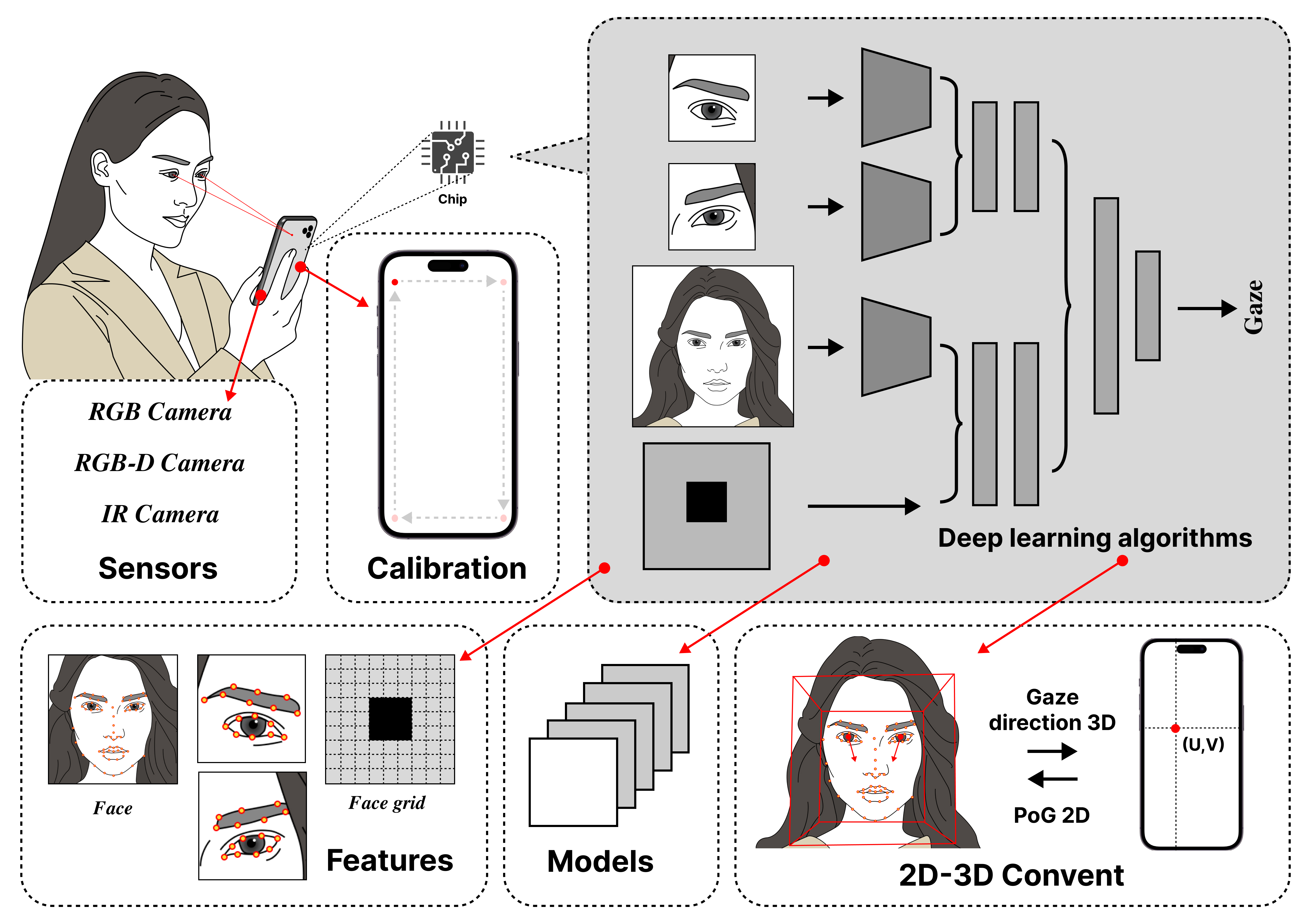}
    \caption{Workflow of gaze estimation based on camera}
    \label{fig:workflow}
\end{figure}
As presented in Figure~\ref{fig:workflow}, the workflow of gaze estimation on handheld mobile devices consists of four main stages: (1) \textit{pre-processing} -- rectifying and processing images or videos acquired from a camera to identify a face and extract features; (2) \textit{learning} -- applying a machine learning technique to estimate a gaze point or vector; (3) \textit{post-processing} -- converting the gaze output to suit the requirements of applications; and (4) \textit{calibration} -- calibrating the gaze estimation model to cater for the characteristics and context of a new environment, device, and user. In the following, we will illustrate each of the above stages.

\subsubsection{Pre-processing and Feature Extraction}\label{subsubsec:preprocessing}
At the pre-processing stage, images are collected from a camera, from which a face can be identified and facial landmarks are extracted. Facial landmarks, referred to as a set of coordinates on an image, are used to locate and represent important regions of the face, including the chin, mouth, nose, eyes, and eyebrows. OpenFace~\cite{baltrusaitis2018openface} and Dlib~\cite{dlib09} are the most commonly used libraries for face identification and landmark extraction~\cite{zhang2015appearance,zhang2017fullface,lian2018multiview}. Recently, researchers are starting to design customised networks for more accurate and richer facial feature extraction, including Multitask Cascaded Convolutional Network (MTCNN) \cite{zhang2016joint}, Deep Alignment Network (DAN) \cite{kowalski2017deep}, Position Map Regression Network (PRNet)~\cite{feng2018joint}, and 3D Dense Face Alignment (3DDFA) \cite{zhu2017face}. Significant effort has also been dedicated to pupil segmentation and detection, including PupilNet~\cite{fuhl2017pupilnet}, Circular Binary Features (CBF)~\cite{10.1145/3204493.3204559} and Boosted-Oriented Edge optimisation (BORE)~\cite{10.1145/3204493.3204558} for real-time pupil detection. 

Once facial landmarks are extracted, face and/or eye images can be cropped. A common way to do so is to crop a square region with the centre of the face and a width. The centre of the face is the averaged coordinate positions of all the facial landmarks and the width can be set as a ratio to the maximum distance between the landmarks; for example, Zhang et al.~\cite{zhang2017fullface} set the ratio as 1.5. As head pose has a significant impact on gaze estimation, often the images need to be rectified, including rotating and shifting to align with a reference point~\cite{cheng2021appearance}. With cropped face and eye images, various features can be extracted, including eye, face, face and eye, and temporal.

\paragraph{Eyes}
The eye is intrinsically connected to gaze estimation, as any change in the gaze direction leads to a corresponding alteration in the eye's appearance. For instance, eye rotation affects the iris's position and the eyelid's shape, resulting in a shift in the gaze direction. This connection enables gaze estimation based on the eye's appearance. Many methods leverage both eyes as cascading features for gaze estimates, with examples including, Minst~\cite{zhang2015appearance}, SAGE~\cite{9021975}, GoogleGaze~\cite{valliappan2020accelerating}, EyeNet~\cite{park2020towards}, and EVE-SCPT~\cite{bao2022individual}. However, for gaze estimation from handheld mobile devices, the visibility of both eyes only has 68.18\%~\cite{khamis2018understanding}, so monocular features remain valuable and are used by DPGE~\cite{park2018deep}, Deng et al.~\cite{zhu2017monocular} and OneEye-Net~\cite{athavale2022oneeye}.

\paragraph{Face}
Facial images provide valuable information on the head pose, which is beneficial for gaze estimation. Numerous methods utilise facial features as input for their models, such as FullFace~\cite{zhang2017fullface}, GazeTR~\cite{cheng2021gaze}, L2CS-Net~\cite{abdelrahman2022l2csnet}, GazeCLR~\cite{jindal2022gazecrl}, SE-Gaze~\cite{o2022self}, FreeGaze~\cite{du2022freegaze}, GazeNAS~\cite{nagpure2023GazeNAS}, and Gaze360~\cite{kellnhofer2019gaze360}. However, facial images can also contain redundant information. Researchers, as seen in PureGaze~\cite{cheng2022puregaze} and other models~\cite{ogusu2019lpm, zhang2017fullface, zhang2020learning}, have attempted to filter out irrelevant features in facial images, enabling the model to focus on the core facial features that are common to all. This optimisation mechanism implicitly eliminates gaze-irrelevant features, thus enhancing the gaze estimation network's robustness. Additionally, facial landmarks are employed as supplementary features to model head pose and eye position.

\paragraph{Face and Eyes}
In order to have robust features, some studies have employed both eye and face features as input to have robust features, with these networks typically being multi-streamed to obtain information from the face and eyes simultaneously. Facial landmarks and face grid are also utilised as additional features to model for head pose, eye position and spatial information, often used in conjunction with eye and face features. Examples of such methods include iTracker~\cite{krafka2016eye}, GazeAttentionNet~\cite{huang2022gazeattentionnet}, Dilated-Net~\cite{chen2018appearance}, TAT~\cite{guo2019generalized}, AFF-Net~\cite{bao2021adaptive}, iMon~\cite{huynh2021imon}, RecurrentGaze~\cite{palmero2018recurrent},GAZEL~\cite{park2021gazel},  RT-Gene~\cite{fischer2018rt}, and HAZE-Net~\cite{yun2022haze}.

\paragraph{Static and Temporal Features}
Features obtained from images are static; however, the information captured by a camera is dynamic, with a correlation between frames. To improve the robustness on prediction, some studies~\cite{palmero2018recurrent, kellnhofer2019gaze360, huynh2021imon, park2020towards} investigate the temporal information by extracting features from video and learning their temporal correlations between frames. Temporal features, such as optical flow~\cite{huynh2021imon} and eye movement dynamics~\cite{wang2019neuro}, have been used to improve gaze estimation accuracy. Optical flow provides information about motion between frames. Gaze360~\cite{kellnhofer2019gaze360}, RecurrentGaze~\cite{palmero2018recurrent}, EyeNet~\cite{park2020towards} and EVE-SCPT~\cite{bao2022individual} directly apply models such as LSTM to obtain correlation information between video frames.

\subsubsection{Learning}\label{subsec:deeplearning}
\begin{figure}[!htbp]
    \centering
    \includegraphics[width=\columnwidth]{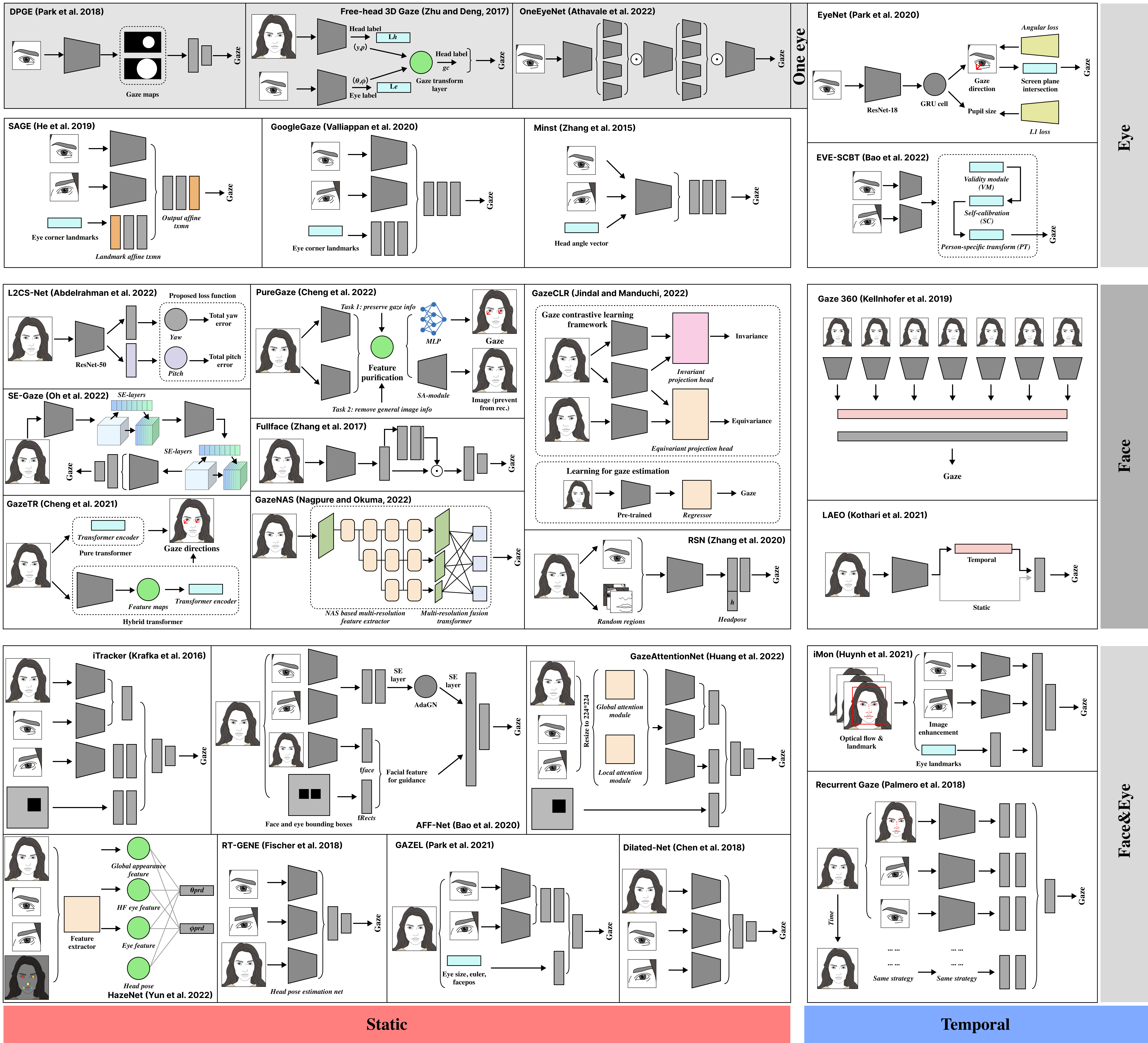}
    \caption{Overview of the deep learning architectures for Gaze Estimation 2D \& 3D tasks}
    \label{fig:architecture}
\end{figure}

Table~\ref{tab:DLModels} and Figure~\ref{fig:architecture} present the mainstream deep learning models applied to gaze estimation, which are built on Convolutional Neural Networks (CNN), Recurrent Neural Networks (RNN), and more recently Transformers. 

\paragraph{Convolutional Neural Networks}
CNN is one of the most commonly used techniques for extracting facial and eye features.  In gaze estimation, VGG~\cite{simonyan2014very,zhang2017mpiigaze,zhang2015appearance,fischer2018rt},  ResNet~\cite{he2016deep,kellnhofer2019gaze360,zhang2020eth}, LeNet~\cite{lecun1998gradient}, AlexNet~\cite{krizhevsky2012imagenet}, MobileNet-V2~\cite{chang2021high} and DenseNet~\cite{huang2017densely} are the mostly adopted state-of-the-art architectures. Researchers also design novel architectures for face and eye feature extraction. Zhang et al.~\cite{zhang2017fullface} have proposed a customised CNN with spatial weights for full-face 2D and 3D gaze estimation. The first few layers of the CNN extract image features, and then the last few layers are dedicated to learning spatial weights for the activation of the last convolutional layer. The purpose of the spatial weights is to indicate the importance of different regions of the face for gaze estimation. 

Features from eye images can be combined with the head pose, and the eye features can be explored independently (with each of the two eyes) or jointly (with two eyes together). Fischer et al.~\cite{fischer2018rt} have applied two VGG-16 networks~\cite{simonyan2014very} to learn individual features from two eye images and then concatenate these features to predict the yaw and pitch gaze angles. Cheng et al.~\cite{cheng2018appearance} have proposed an asymmetric regression evaluation network where a four-stream CNN is applied to extract and combine features on individual eye images, which are used in linear regression to estimate the 3D gaze direction. 

\paragraph{Recurrent Neural Networks}
RNN~\cite{rumelhart1986learning} and Long short-term memory (LSTM)~\cite{hochreiter1997long}  have been employed to explore and leverage the temporal relationships between consecutive frames to improve the estimation accuracy. 
Palmero et al.~\cite{palmero2018recurrent} have proposed a Recurrent CNN Regression Network for 3D gaze estimation. The network is composed of 3 modules: individual, fusion, and temporal. The individual module learns features from each appearance cue independently and consists of a two-stream CNN: one for processing face images and the other for jointly learning eye images. The fusion module combines the extracted features of each appearance stream in a single vector along with the normalised landmark coordinates. Then, it learns a joint representation between modalities in a late-fusion fashion. Both individual and fusion modules are applied to each frame, and the resulting feature vectors of each frame are input to the temporal module based on a many-to-one recurrent network. This module leverages sequential information to predict the normalised 3D gaze angles of the last frame of the sequence using a linear regression layer.

Bidirectional LSTMs are another popular choice for modelling temporal relationships between successive frames~\cite{zhou2020improved,kellnhofer2019gaze360}. Kellnhofer et al.~\cite{kellnhofer2019gaze360} have utilised 7 consecutive sequence frames to obtain continuous head and eye movements in the video stream to improve the accuracy of predicting the gaze of the central frame in the input sequence. Specifically, a head crop from each frame is individually processed by ResNet-18, and the extracted features are fed to a bidirectional LSTM with two layers that take the sequence of forward and backward vectors. Finally, these vectors are concatenated and passed through a fully connected layer to produce two outputs: gaze prediction and error quantile estimation. 

Combining the gaze direction with the optical flow in the eye region is another way of exploiting temporal information in motion. Bace et al.~\cite{bace2020combining} have combined gaze direction with the optical flow in the eye region to identify the target that a user is following. The system consists of two components: one for taking a facial image as input and predicting 2D gaze points, and the other for estimating the motion pattern between consecutive frames. Aggregating the outputs from these two components can improve the robustness of estimation.

\paragraph{Transformers}
Transformer, originating from natural language processing~\cite{vaswani2017vit}, contains self-attention layers, layer normalisation and multi-layer perception layers. Compared to RNN, self-attention layers have global computations and better memory to process long-sequence tasks. Transformers also perform well in computer vision tasks by replacing words in NLP tasks with non-overlapping image patches, called vision transformer (ViT)~\cite{dosovitskiy2021vit}. 
GazeTR~\cite{cheng2021gaze} is an early ViT adopter in the field of gaze estimation. They apply the original ViT architecture to gaze estimation and also propose a hybrid ViT architecture combined with a CNN feature extractor. The hybrid ViT architecture achieved better results than ViT alone on various datasets. GazeNAS~\cite{nagpure2023GazeNAS} is another early adaptor of ViT, where they propose a light-weighted ViT that only has 1 million parameters and uses 0.28 GFLOPs. It involves neural architecture search (NAS) from reinforcement learning as a multi-resolution feature extractor. GazeNAS achieves state-of-the-art results on various benchmark datasets.

\paragraph{Semi-supervised and Unsupervised Learning}
Most deep learning algorithms require a large number of training data and collecting high-quality gaze datasets is a time- and effort-consuming task. Therefore, in recent years, we have witnessed more and more semi-supervised and unsupervised learning techniques being developed to reduce the reliance on labelled data. 
Kothari et al.~\cite{kothari2021weakly} have proposed to curate videos from the Web where people are ``looking at each other" (LAEO), and annotate each frame with whether LAEO is present. They design a weakly supervised algorithm for learning 3D gaze information by enforcing scene-level geometric 3D and 2D LAEO constraints between pairs of faces. Dubey et al.~\cite{dubey2019unsupervised} propose an unsupervised learning technique based on a large in-the-wild dataset that contains many facial images from the Web. They localise the pupil-centre of each eye and use them to determine the region in which the subject is looking. Similarly, Yu et al.~\cite{yu2020unsupervised} also propose to learn gaze representation from unannotated eye images and then use a few labelled calibration samples for gaze estimation.

\begin{table}
\centering
\caption{A list of representative appearance-based gaze estimation models\label{tab:DLModels}}
\begin{tabular}{l|llllccc}
\hline 
\multirow{2}{*}{\diagbox{Methods}{Attributes}} & \multirow{2}{*}{Year} & \multicolumn{1}{c}{\multirow{2}{*}{Feature}} & \multicolumn{1}{c}{\multirow{2}{*}{Model}} & \multicolumn{1}{c}{\multirow{2}{*}{Dataset}} & \multicolumn{2}{c}{PoG(cm)} & \multirow{2}{*}{Direction($^{\circ}$)} \\
 &  & \multicolumn{1}{c}{} & \multicolumn{1}{c}{} & \multicolumn{1}{c}{} & Tablet & Phone &  \\
\hline
\hline 
Minst~\cite{zhang2015appearance} & 2015 & Eyes, Lmks & CNN & MPIIGaze & - & - &  6.27$^{\circ}$ \\
iTracker~\cite{krafka2016eye} & 2016 & Face, Eyes, Grid & CNN & GazeCapture & 2.81 & 1.86 & - \\
FullFace~\cite{zhang2017fullface} & 2017 & Face & CNN & MPIIFaceGaze & - & - & 4.80$^{\circ}$ \\
Dilated-Net~\cite{chen2018appearance} & 2018 & Face, Eyes & CNN & MPIIGaze & - & - &  5.12$^{\circ}$ \\
RT-Gene~\cite{fischer2018rt} & 2018 & Face, Eyes & CNN & MPIIGaze & - & - & 4.66$^{\circ}$ \\
DPGE~\cite{park2018deep} & 2018 & Eye & CNN & MPIIGaze & - & - &  4.50$^{\circ}$ \\
RecurrentGaze~\cite{palmero2018recurrent} & 2018 & Face, Eyes, Lmks & CNN+LSTM & EyeDiap & - & - & 3.38$^{\circ}$  \\
Gaze360~\cite{kellnhofer2019gaze360} & 2019 & Face & CNN+LSTM & Gaze360 & - & - & 11.1$^{\circ}$ \\
SAGE~\cite{9021975} & 2019 & Eyes, Lmks & CNN & GazeCapture & 2.72 & 1.78 & - \\
TAT~\cite{guo2019generalized} & 2019 & Face, Eyes & CNN & GazeCapture & 2.66 & 1.77 & - \\
RSN~\cite{zhang2020learning} & 2020 & Face & CNN & MPIIGaze & - & - & 4.50$^{\circ}$  \\
GoogleGaze~\cite{valliappan2020accelerating}  & 2020 & Eyes, Lmks & CNN & GazeCapture & - & 1.92 & - \\
EyeNet~\cite{park2020towards} & 2020 & Eyes & CNN+GRU & EVE & \multicolumn{2}{c}{3.85} & 3.48$^{\circ}$ \\
AFF-Net~\cite{bao2021adaptive} & 2021 & Face, Eyes & CNN & GazeCapture & 2.30 & 1.62  & - \\
iMon~\cite{huynh2021imon} & 2021 & Face, Eyes & CNN & GazeCapture & 1.94 & 1.49 & - \\
GAZEL~\cite{park2021gazel} & 2021 & Face, Eyes, Lmks & CNN & Private dataset & 2.91 & - & - \\
PureGaze~\cite{cheng2022puregaze} & 2021 & Face & CNN+SA+MLP & ETH-XGaze & - & - & 4.50$^{\circ}$ \\
GazeTR~\cite{cheng2021gaze} & 2021 & Face & ViT & MPIIFaceGaze & - & - & 4.00$^{\circ}$ \\
EVE-SCPT~\cite{bao2022individual} & 2022 & Eyes & CNN+GRU & EVE & \multicolumn{2}{c}{2.75} & 2.49$^{\circ}$ \\
GazeAttentionNet~\cite{huang2022gazeattentionnet} & 2022 & Face, Eyes, Grid & CNN+MLP & ETH-XGaze & - & - & 4.5$^{\circ}$ \\
L2CS-Net~\cite{abdelrahman2022l2csnet} & 2022 & Face & CNN & Gaze360 & - & - & 9.02$^{\circ}$ \\
GazeCLR~\cite{jindal2022gazecrl} & 2022 & Face & CNN & EVE & - & - & 4.15$^{\circ}$ \\
OneEye-Net~\cite{athavale2022oneeye} & 2022 & Eye & CNN & GazeCapture & \multicolumn{2}{c}{2.31} & - \\
SE-Gaze~\cite{o2022self} & 2022 & Face & CNN+SE+MLP & MPIIFaceGaze & - & - & 4.04$^{\circ}$ \\
FreeGaze~\cite{du2022freegaze} & 2022 & Face & CNN & ETH-XGaze & - & - & 2.95$^{\circ}$ \\
HAZE-Net~\cite{yun2022haze} & 2022 & Face, Eyes, Lmks & CNN & EyeDiap & - & - & 4.12$^{\circ}$ \\
GazeNAS~\cite{nagpure2023GazeNAS} & 2023 & Face & ViT & MPIIFaceGaze & - & - & 3.96$^{\circ}$ \\
\hline
\end{tabular}
\end{table}

\subsubsection{Post-processing}
Depending on the available data for training gaze estimation algorithms and the requirements of the applications, there is a need to convert between 2D and 3D gaze. The conversion is performed by rotating and translating the camera coordinate system (CCS) and screen coordinate system (SCS)~\cite{cheng2021appearance}. To convert a 2D gaze point to a 3D gaze direction, we first obtain the rotation $R_s$ and translation $T_s$ matrices of SCS with respect to CCS by geometric calibration. With these two matrices, we can compute the 3D gaze target with $t = R_s[u,v,0]^T + T_s$, which is the intersection of gaze direction and the screen. The target $t$ will be used to derive the 3D gaze direction $g = k(t-o)$, where $k$ is the factor $\frac{1}{||t-o||}$ and $o$ is the gaze origin; for example, the face or eye centre. 

To convert a 3D gaze direction to a 2D point, we will still require $R_s$ and $T_s$ and the gaze origin and then revert the calculation process. First, we need to calculate the 3D gaze target vector $t=(x,y,z)$ which is the interaction of gaze direction to the screen. To do so, we use two equations: the line of sight and the screen plan. Given the origin $o=(o_x, o_y, o_z)$ and a 3D gaze direction $g=(g_x, g_y, g_z)$, we can get the equation of the line of sight as
\begin{equation}\label{eq:line_of_sight}
    \frac{x-o_x}{g_x} = \frac{y-o_y}{g_y} = \frac{z-o_z}{g_z}.
\end{equation}
The equation of the screen plane is the relation between the target vector $t$ and the rotation matrix $R_s$ and the translation matrix $T_s$. From $R_s$, we can derive a normal vector of screen plane $n=R_s[:,2]=(n_x,n_y, n_z)$. Given $T_s=[t_x, t_y, t_z]^T$, we can deduce the equation of screen plane as
\begin{equation}\label{eq:screen_plane}
    n_xx + n_y y + n_z z = n_x t_x + n_y t_y + n_z t_z. 
\end{equation}

Solving the above equations \ref{eq:line_of_sight} and \ref{eq:screen_plane} gives us the target vector $t$. Then we can compute the corresponding 2D gaze point as $p=(u, v, 0) = R^{-1}_s (t - T_s)$.

With handheld mobile devices, most of the existing interactive systems use 2D gaze points. The reason is that the holding posture of devices often changes, and the relation between the eye and the screen is not stable, which either leads to inaccurate calculation of the direction or requires constant recalibration of the rotation and translation matrices. Nowadays, research effort is increasingly devoted to 3D gaze direction; for example, Zhang et al. have set up a multi-camera system to capture over 500 gaze directions with various illumination conditions~\cite{zhang2020eth}. 3D gaze directions depending on the coordinate system of the head and face may improve robustness for mobile devices in complex environments.

\subsubsection{Calibration}\label{subsec:calibration}
Eye-tracking systems often need to be calibrated; otherwise, the gaze output might incur too large errors to be useful. Calibration is the process of adjusting and customising the gaze output to reflect the spatial geometry of the camera, the screen, and personal difference~\cite{Drewes2019timecalibration} in order to improve the estimation accuracy. The calibration process consists of \textit{data collection} and \textit{training}. 

\paragraph{Data collection.} The common way to collect ground truth data for calibration is to design an interactive interface to guide a user's gaze attention. The point-based method is the most used example, where users are asked to fixate at a point for a few seconds. There are often between 5 and 16 points being displayed consecutively at various locations on a screen~\cite{Drewes2019timecalibration}. Another popular method is pursuit-based, where users are asked to follow their gaze on a moving target, and the trajectory of the movement can be any shape such as a circle~\cite{Drewes2019timecalibration} or a rectangle~\cite{lei2023DynamicRead}. Compared to the point-based method, the pursuit-based method requires less time to collect the same number of data points and provides a better user experience. 

The above data collection is considered as \textit{explicit}, requiring the users' attention and voluntary actions. In contrast, \textit{implicit} calibration collects ground-truth data from the background by estimating user attention in possible fixation locations; for example, mouse and keyboard events~\cite{liebling2014gaze, kasprowski2016implicit, huang2016autocalibration}, typing on the on-screen keyboard and screen touch events~\cite{jiang2020howtype, weill2016you}. A recent study~\cite{yang2021vgaze} employs visual saliency, which has distinctive perceptual properties from their surroundings that attract users' gaze attention~\cite{amso2014eye, koch1987shifts}. For example, independent continuous objects such as a sailing speedboat and a bright moon in the dark in the video are used as visually salient locations for collecting calibration points implicitly~\cite{yang2021vgaze}. 

\paragraph{Calibrator Training.} 
Table~\ref{tab:calibrationmethods} presents a list of calibration techniques. Earlier calibrators are often adaptive linear regression techniques~\cite{lu2014adaptive}. Recently, domain adaptation techniques have been applied to tackle the personalised calibration problem. For example, Cui et al.~\cite{cui2017specialized} have applied Geodesic Flow Kernel (GFK) to adapt the gaze estimator trained on adult data (the source domain) to predict gaze on children (the target domain). 

With deep learning, transfer learning has been widely attempted; that is, either fine-tuning the fully connected layer from a pre-trained CNN model~\cite{zhang2018PersonSpecific} or taking the features extracted on a CNN to train a Support Vector Regression (SVR)~\cite{krafka2016eye, valliappan2020accelerating, lei2023DynamicRead, li2020evaluation}. These techniques are sensitive to environmental changes~\cite{lu2014adaptive,krafka2016eye}. To overcome this problem, Wang et al.~\cite{wang2019generalizing} introduce a Bayesian adversarial learning technique in which an adversarial learning block is employed to learn generalisable gaze features to various appearances and head poses.

The FAZE project~\cite{park2019few} applies a few-shot learning technique, Model-Agnostic Meta-Learning (MAML), to tailor a gaze estimation model to an individual only with a few calibration samples. It first learns robust features via an encoder-decoder architecture, which captures latent features on head orientation, gaze direction, and the appearance around the eye regions. Then it fine-tunes the model with a few calibration samples on individuals. Often the size of these samples is small and thus leads to an overfitting problem. To resolve this problem, MAML is adopted, which employs 2-step gradient updates to learn the optimal weights on the person-specific models.  It has the advantage of minimising the generalisation loss of the network~\cite{park2019few}. 

To calibrate the estimation model for individual users, there are efforts from the interaction perspective; that is, using the relevant eye movements generated during the interaction for calibration~\cite{kasprowski2016implicit} and accuracy maintenance~\cite{huang2019saccalib}. We will discuss this practice in Section~\ref{sec:future}.

\begin{table}
\centering
\caption{A List of Calibration Methods on Handheld Mobile Devices}
\label{tab:calibrationmethods}
\begin{tabular}{l|ccccccc} 
\hline
\multirow{2}{*}{\diagbox{Project}{Attributes}} & \multirow{2}{*}{Year} & \multicolumn{2}{c}{Data Collection} & \multicolumn{1}{c}{\multirow{2}{*}{Calibrator}} & \multicolumn{2}{c}{PoG(cm)} & \multirow{2}{*}{Direction($^{\circ}$)} \\
 &  & Explicit & Implicit & \multicolumn{1}{c}{} & Tablet & Phone &  \\ 
\hline
\hline
iTracker~\cite{krafka2016eye} & 2016 & 13-Point & - & SVR & 2.12 & 1.34 & - \\
FAZE~\cite{park2019few} & 2019 & - & - & MAML & - & - & 3.08$^{\circ}$ \\
GoogleGaze~\cite{valliappan2020accelerating} & 2020 & Points & - & SVR & - & 0.46 & - \\
GazeRefineNet~\cite{park2020towards} & 2020 & - & Visual Saliency & -* & \multicolumn{2}{c}{2.75} & 2.49$^{\circ}$ \\
GazeL~\cite{park2021gazel} & 2021 & - & Touch Event & SVR & - & 1.58 & - \\
vGaze~\cite{yang2021vgaze} & 2021 & - & Visual Saliency & Clustering + LR & - & 1.51 & - \\
DAGE~\cite{li2021device} & 2021 & 9-Point & - & MLP & 2.43 & 1.58 & - \\
iMon~\cite{huynh2021imon} & 2021 & 5-Point & - & Kappa angle+LR & 1.59 & 1.11 & - \\
EasyGaze~\cite{cheng2022easygaze} & 2022 & 9-Point & - & LR & - & - & 1.93$^{\circ}$ \\
DynamicRead~\cite{lei2023DynamicRead} & 2023 & Pursuit & - & SVR & - & 0.95 & - \\
\hline
\end{tabular}
\begin{tablenotes}
      \small
      \item Note: SVR - Support Vector Regression; MAML - Model-Agnostic Meta-Learning; MLP - Multi Layer Perceptron; LR - Liner Regression without a specific model; * - GazeRefineNet is a label-free PoG refinement model that employed visual saliency.
    \end{tablenotes}
\end{table}

\subsection{Summary}
From the analysis in Section~\ref{subsec:sensors}, we observe that mobile devices are increasingly integrating advanced cameras, such as RGB-D and IR cameras, facilitating the development of appearance-based gaze estimation on mobile devices. As shown in Table~\ref{tab:DLModels}, numerous deep learning techniques have been employed to enhance gaze estimation precision, with the input for gaze estimation evolving from facial landmarks to eye landmarks, and eventually to original face and eye images. The evolution is driven by the challenge of changing holding posture of devices and rotation of heads. For example, full-face images have been used to detect head orientation, eye position, eyelid openness, and eyebrow movement. Such information can be used in conjunction with features extracted from face images. In addition, gaze estimation is highly correlated with eye appearance. Subtle muscle changes in the eye area can lead to a change in gaze direction. Intuitively, the up-and-down movement of the eye drives the interplay of surrounding muscles, such as the eyelids and iris. Therefore, eye appearance features from the whole frame are increasingly employed in gaze estimation. 

Most of deep learning models are built on state-of-the-art CNNs or custom CNNs, while recurrent models and ViT have not demonstrated significant improvements over CNNs. As detailed in Tables \ref{tab:DLModels} and \ref{tab:calibrationmethods}, the accuracy of gaze estimation can reach 1.49 cm on mobile phones, this level can support coarse-grained gaze interactions on handheld mobile devices, considering that widget sizes typically range from 0.9 cm to 1.2 cm. With calibration, the precision can be further improved; for example, Lei et al. have applied SVR for pursuit calibration and reduced their errors to 0.95cm, which enables various types of gaze interfaces to support scrolling actions in a reading application~\cite{lei2023DynamicRead}. Much of the current research prioritises enhancing gaze estimation accuracy in static conditions and fixed postures. Future research should focus on achieving high-accuracy eye-tracking in dynamic real-world scenarios, where users interact with their devices in various natural postures~\cite{lei2023DynamicRead,namnakani2023comparing}.

\section{Datasets}\label{sec:datasets}
\begin{figure}[!htbp]
    \centering
    \includegraphics[width=0.95\textwidth]{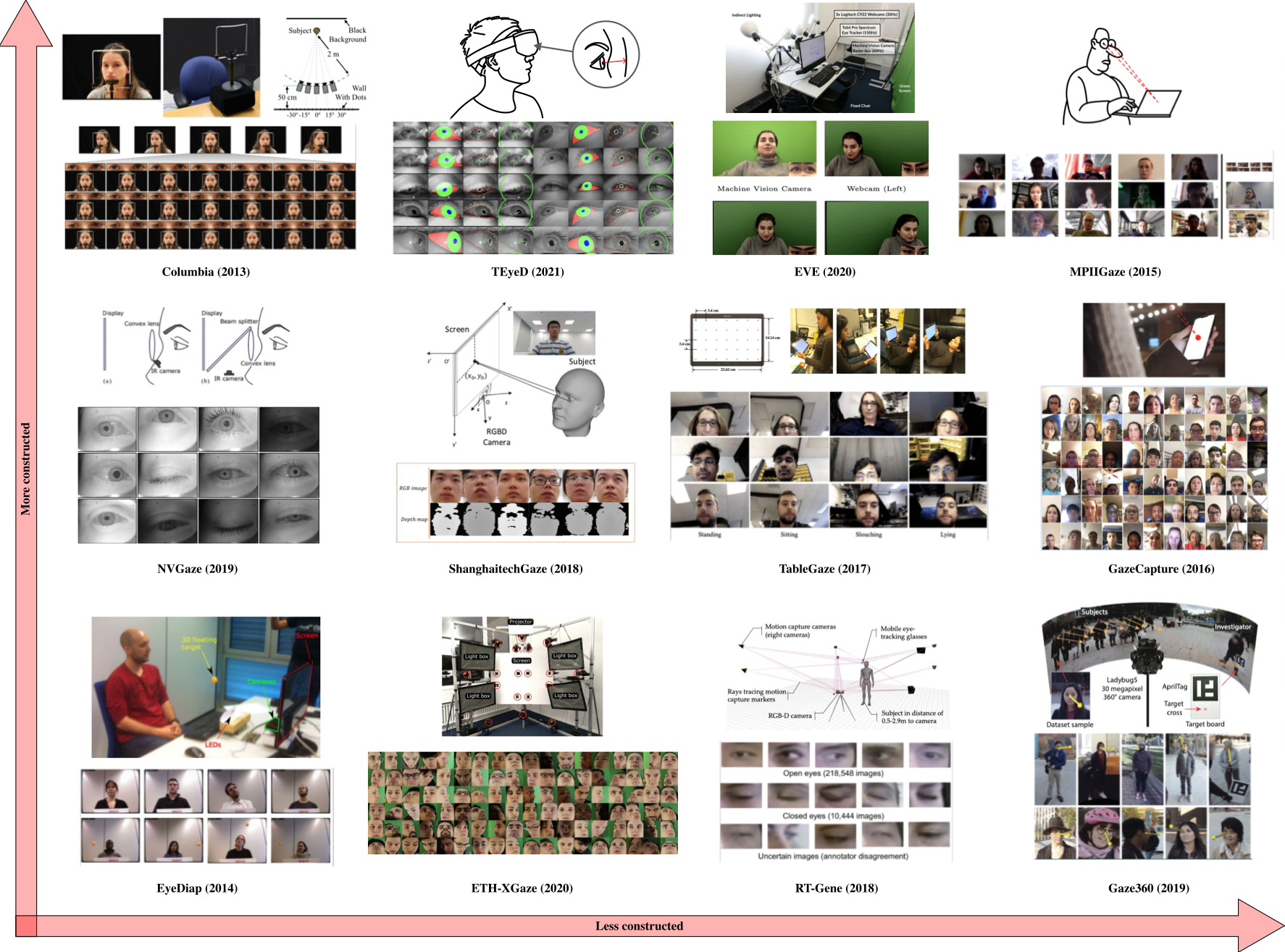}
\caption{Gaze dataset examples and their collection settings, including Columbia~\cite{smith2013gaze}, EyeDiap~\cite{funes2014eyediap}, MPIIGaze~\cite{zhang2015appearance}, GazeCapture~\cite{krafka2016eye}, TableGaze~\cite{huang2017tabletgaze}, ShanghaiTechGaze+~\cite{lian2019rgbd}, RT-GENE~\cite{cortacero2019rt}, Gaze360~\cite{kellnhofer2019gaze360}, NVGaze~\cite{kim2019nvgaze}, EVE~\cite{park2020towards}, and ETH-XGaze~\cite{zhang2020eth}}
    \label{fig:datasets}
\end{figure}

Datasets are an important research aspect in gaze estimation, and we have listed and reviewed all the publicly available gaze datasets in Table~\ref{tab:datasets}. Most of these datasets are collected on the RGB cameras alone (17 out of 24), 3 of them are on RGB-D, and 4 are on IR. There is a trend of moving the platform and collection conditions from more controlled, desktop-based environments towards unconstrained, mobile device-based settings. In the early stage, gaze data is often collected in the laboratory environment and subjects are required to ﬁx their head on a chin rest~\cite{funes2014eyediap, zhang2020eth}. Figure~\ref{fig:datasets} has shown the collection settings and samples of the gaze datasets. More recently, research attention is gradually shifting to real-world settings with a variety of distances to screens, head poses, and illumination conditions. For example, MPIIGaze~\cite{zhang2017mpiigaze} consists of over 213,000 images from 15 people looking at different gaze positions. The dataset is collected over three months during daily laptop usage. GazeCapture~\cite{krafka2016eye} is one of the pioneers in collecting gaze data in in-the-wild environments, containing over 1.4 million images from over 1400 users. TEyeD~\cite{fuhl2021teyed} is the currently largest public dataset of eye images, which contains over 20 million real-world eye images with gaze vector, eye movement types, and pupil and eyelid. The data is collected when users wearing head-mounted eye trackers are performing various tasks, including outdoor sports, daily indoor activities, and car riding. 

We have included the datasets collected on both desktop and mobile devices, even though our focus is on mobile device-based gaze estimation. The reason is that the desktop-based datasets can be used to train a deep learning model for extracting face and eye features. In addition, other benchmark datasets can also be used to train a deep-learning model. For example, the pupil and eyelid datasets~\cite{fuhl2016else,fuhl2015excuse}, face image datasets including VGG-Face~\cite{parkhi2015deep}, Labeled Faces in the Wild dataset (LFW)~\cite{huang2008labeled} and YouTube Faces (YTF)~\cite{wolf2011face}, Celebfaces~\cite{liu2015faceattributes}, BayesianFace~\cite{lu2014learning} can be used to extract facial features, and thus for gaze estimation~\cite{cai2015person}.  

\begin{table}
\centering
\caption{A summary of publicly available gaze datasets}
\label{tab:datasets}
\begin{tabular}{l|cccccccll}
\hline
\multicolumn{1}{c|}{Dataset} & Year & Camera & Gaze & Head Move & IC & Distance & Sub & \multicolumn{1}{c}{Resolution} & \multicolumn{1}{c}{Images} \\ 
\hline
\hline
Columbia \cite{smith2013gaze} & 2013 & RGB  & 2D, 3D  & 5 & 1 & 200cm & 56 & 5184×3456 & 5,880 \\
UT Multiview \cite{sugano2014learning} & 2014 & RGB  & 2D, 3D   & 8  & 1 & 60cm & 50 & 1280×1024  & 64,000 \\
EyeDiap \cite{funes2014eyediap} & 2014 & RGB-D  & 2D, 3D   & C & 2 & 80-120cm & 16 & 1920×1080 & 62,500 \\
OMEG \cite{he2015omeg} & 2015 & RGB  & 3D  & 3 + C & 10 & varying & 50 & 1280×1024 & 44,827 \\
SynthesEyes \cite{wood2015rendering} & 2015 & RGB  & 3D   & C & 4 & varying & 10 & 120×80 & 11,382\\
MPIIGaze \cite{zhang2015appearance} & 2015 & RGB  & 2D, 3D   & C & D & 40-60cm & 15 & 1280×720 & 213,659 \\
GazeFollow \cite{recasens2015they} & 2015 & RGB  & 3D   & C & D & varying & 130,339 & variable & 122,143 \\
GazeCapture \cite{krafka2016eye} & 2016 & RGB  & 2D   & C & D & varying & 1474 & 640×480 & 2,445,504 \\
UnitEyes \cite{wood2016learning} & 2016 & RGB  & 3D   & C & D & 0.5-3cm & N/A & 400 × 300 & 1,000,000 \\
MPIIGazeFace \cite{zhang2017fullface} & 2017 & RGB  & 2D, 3D   & C & D & varying & 15 & 1280×720 & 37,639 \\
TabletGaze \cite{huang2017tabletgaze} & 2017 & RGB  & 2D   & C & 1 & 30-50cm & 51 & 1280 × 720 & 1,785 \\
InvisibleEye \cite{tonsen2017invisibleeye} & 2017 & RGB  & 2D   & N/A & 1 & 0.5-2cm & 17 & 5 × 5 & 280,000 \\
RT-GENE \cite{cortacero2019rt} & 2018 & RGB-D  & 3D   & C & 1 & 80-280cm & 15 & 1920×1080 & 122,531 \\
Gaze360 \cite{kellnhofer2019gaze360} & 2019 & RGB  & 3D   & C & D & varying & 238 & 4096×3382 & 172,000 \\
NVGaze \cite{kim2019nvgaze} & 2019 & IR  & 2D  & 1 & 1 & 0.5-3cm & 30 & 1280*960 & 4,500,000 \\
SHTechGaze \cite{lian2018multiview} & 2018 & RGB  & 2D & C & D & varying & 137 & 1920×1080 & 233,796 \\
SHTechGaze+ \cite{lian2019rgbd} & 2019 & RGB-D  & 2D   & C & D & varying & 218 & 1920×1080 & 165,231\\
EVE \cite{park2020towards} & 2020 & RGB  & 2D, 3D   & C & 1 & varying & 54 & 1920×1080 & 12,308,334 \\
ETH-XGaze \cite{zhang2020eth} & 2020 & RGB  & 3D   & C & 16 & 100cm & 110 & 6000×4000 & 1,083,492\\
GW \cite{kothari2020gaze} & 2020 & IR  & 3D   & C & D & 0.5-3cm & 19 & 1920×1080 & 5,800,000\\
LAEO \cite{kothari2021weakly} & 2021 & RGB & 3D & C & D & varying & 485 & variable & 800,000 \\       
GOO \cite{tomas2021goo} & 2021 & RGB  & 3D   & C & D & varying & 100 & variable & 201,552 \\
OpenNEEDS \cite{emery2021openneeds} & 2021 & IR  & 3D   & C & 1 & 0.5-3cm & 54 & 128×71 & 2,086,507 \\
TEyeD \cite{fuhl2021teyed} & 2021 & IR  & 2D, 3D & C & 1 & 0.5-3cm & 132 & variable & 20,867,073 \\
\hline
\end{tabular}
\begin{tablenotes}
      \small
      \item Note: The columns of the above table are: (1) the publication \textit{year}; (2) the type of \textit{camera}; (3) type of \textit{gaze} (2D PoG or 3D direction); (4) the number of variations in \textit{Head Move} in reference to the screen,  and \textit{C} refers to continuous head movements; (5)  illumination conditions (\textit{IC}) where a number in this column refers to the number of illumination conditions being considered, \textit{D} refers to the ambient light in normal daytime conditions; (6) the \textit{distance} to the camera(s); (7)  the number of \textit{Sub}jects; (8) the resolution of each image; and (9) the number of \textit{images}. 
    \end{tablenotes}
\end{table}

\section{Gaze Analytics}\label{sec:gazeanalysis}
To design gaze-based applications, we often need to process gaze and analyse them into high-level eye movement patterns. This is currently under investigated in the handheld gaze estimation research area. To bridge the gap, this section will first introduce eye physiology to provide a background on the meaning and function of eye movements, and then illustrate how to process gaze data and define high-level gaze events and eye movement patterns. 

\begin{figure}[!htbp]
    \centering
    \includegraphics[width=\columnwidth]{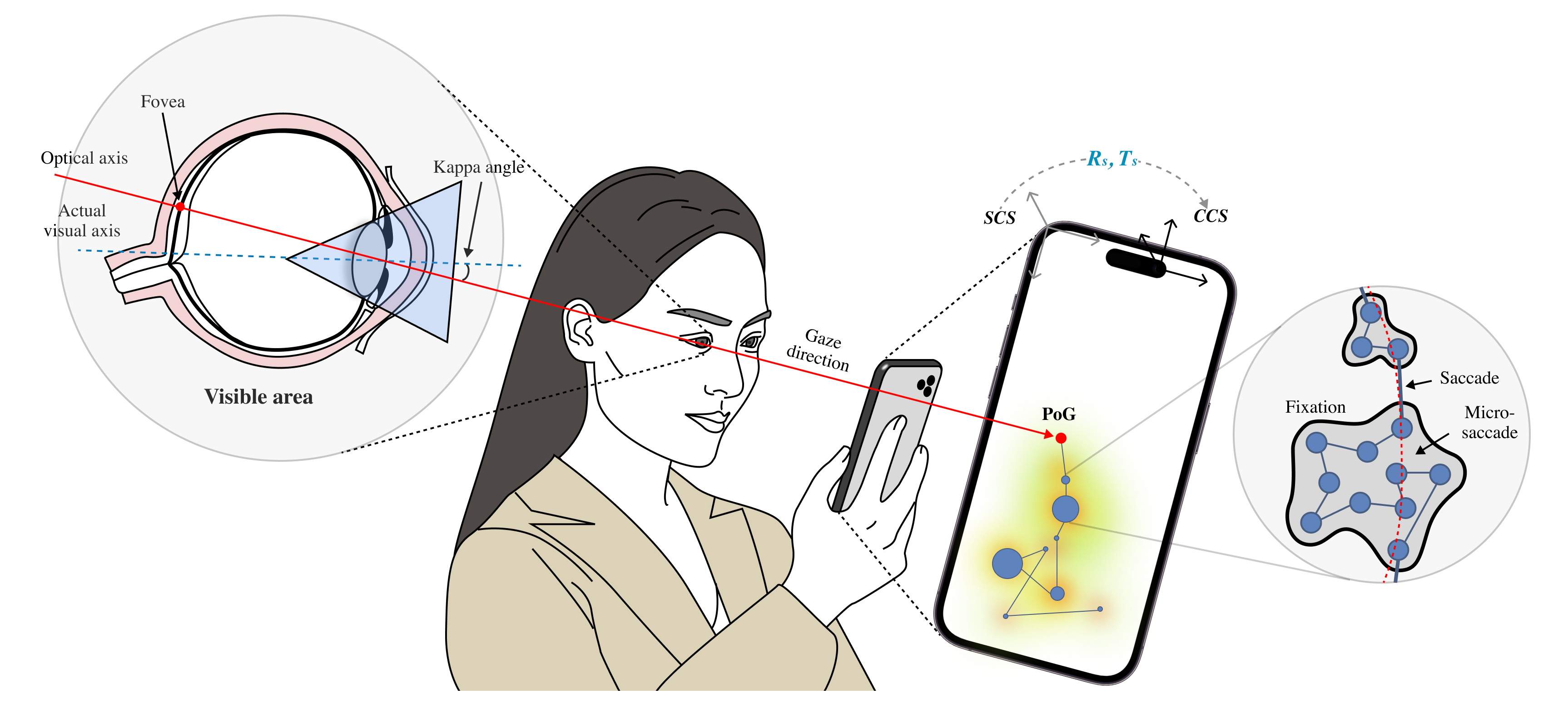}
    \caption{From eye physiology to basic gaze events}
    \label{fig:eyemovement}
\end{figure}

\subsection{Understanding of Eye Movement}\label{subsec:eyemovement}

Eye physiology plays a crucial role in gaze estimation and guides the development of gaze-interactive applications. Grasping the principles of eye movements and their relation to human consciousness levels is essential when designing ergonomic applications. Humans gather information about the external environment through their eyes, which involves continuous voluntary or involuntary movements, enabling the eyes to obtain a steady and continuous visual stimulus. These eye movements can be categorised into various types, such as fixation and saccades, as illustrated in Figure~\ref{fig:eyemovement} 

\begin{enumerate}
    \item \textit{Microscopic eye movements} encompass tremor, microsaccade, and drift, which underlie eye movements like fixation, saccade, and smooth pursuit~\cite{martinez2004role}. Tremor is a periodic, wave-like eye movement with a frequency of 90-105Hz and is the smallest eye movement. It has potential for visual perception~\cite{martinez2004role, martinez2013impact, alexander2018microsaccade}. Microsaccades are small, fast, jerk-like eye movements that occur during voluntary fixational eye movements~\cite{martinez2004role, rolfs2009microsaccades, ryan2019microsaccades}. They can serve as a signal for fixation detection algorithms~\cite{engbert2003microsaccades} and rotate around the point of fixation with small amplitudes~\cite{engbert2003microsaccades}. Drift and tremor typically occur simultaneously and between microsaccades.
    
    \item \textit{Fixation} stabilises the retina on a stationary object of interest to precept and process detailed information from the focused area. This eye movement is essential for tasks that require high visual acuity, such as reading or observing fine details. During fixation, microsaccades occur 1-2 times per second, it help to prevent the fading of the retinal image and maintain visual perception during fixation.~\cite{engbert2003microsaccades, martinez2013impact}.
    
    \item \textit{Saccades} are rapid, ballistic eye movements that occur between fixations, enabling the eye's fovea to continuously locate and track new objects in the visual filed~\cite{duchowski2017eye}. As the fastest eye movements, saccades have speeds ranging from 30 to 900 degrees per second, influenced by factors such as target distance, amplitude, and individual differences. Saccades play a crucial role in tasks like reading, environmental scanning, and searching objects of interest~\cite{ryan2019microsaccades, duchowski2017eye}.
    
    \item \textit{Smooth Pursuit} is the action of stabilising the gaze on a moving visual target. Smooth pursuit has three phases: initiation, maintenance, and termination~\cite{fukushima2013cognitive}. In the initiation phase, a delay of 100-130ms occurs between the target's movement and the start of smooth pursuit, with the first 100ms of smooth pursuit being in the open-loop phase~\cite{buonocore2019eye}. It generally reaches peak eye velocity of 30$^{\circ}$ per second within 220-330ms after the response onset at the target. During maintenance, the eye may exhibit 3-4Hz oscillations for corrective shifts to realign the target image at the fovea. When the target stops, the smooth pursuit movement typically ends within 100ms~\cite{robinson1986model, fukushima2013cognitive}. The human eye can use numerous available signals to provide cues and predictions for future smooth pursuit movements; for example, eyes continue to pursue the target after it disappears or during occlusion~\cite{kowler2019predictive}. Intuitively, the velocity of smooth pursuit lies between that of saccades and fixation.
    
    \item \textit{Blinking} protects the eyes by spreading tears to the corneal surface and blinking periodically keeps the cornea moist. Fixation and blink frequency can be affected by external factors such as humidity and illumination condition as well as internal factors such as cognitive load and fatigue~\cite{martinez2004role}. 
\end{enumerate}

In human visual systems, these eye movements are driven by the attention mechanisms: bottom-up and top-down, which have formed the basis for different gaze interactive applications. The former mechanism refers to the fact that the fixation point of the eye may be changed by external stimuli that direct attention to discriminative areas of the scene. The latter mechanism is driven by internal stimuli of cognition, which often reflect users' intentions. It makes information available in working memory, and people will consciously pay attention to the scene area that is important to the current behavioural target or task~\cite{awh2006interactions,barz2020visual}. Sattar et al.~\cite{sattar2015prediction} learn the compatibility between user fixation scan path and potential targets to predict the correct target image. Wang et al.~\cite{wang2021scanpath} design a model that learns visual saliency and information visualisation of scan paths based on a sequence of eye movements. These explorations based on gaze and eye movements have great potential for predictive reasoning about users' intentions and interests. The visual attention mechanisms have formed the foundation of these gaze interactions. 

\subsection{Gaze Data Analytics}\label{subsec:eyedata}

\begin{figure}[!htbp]
    \centering
    \includegraphics[width=\columnwidth]{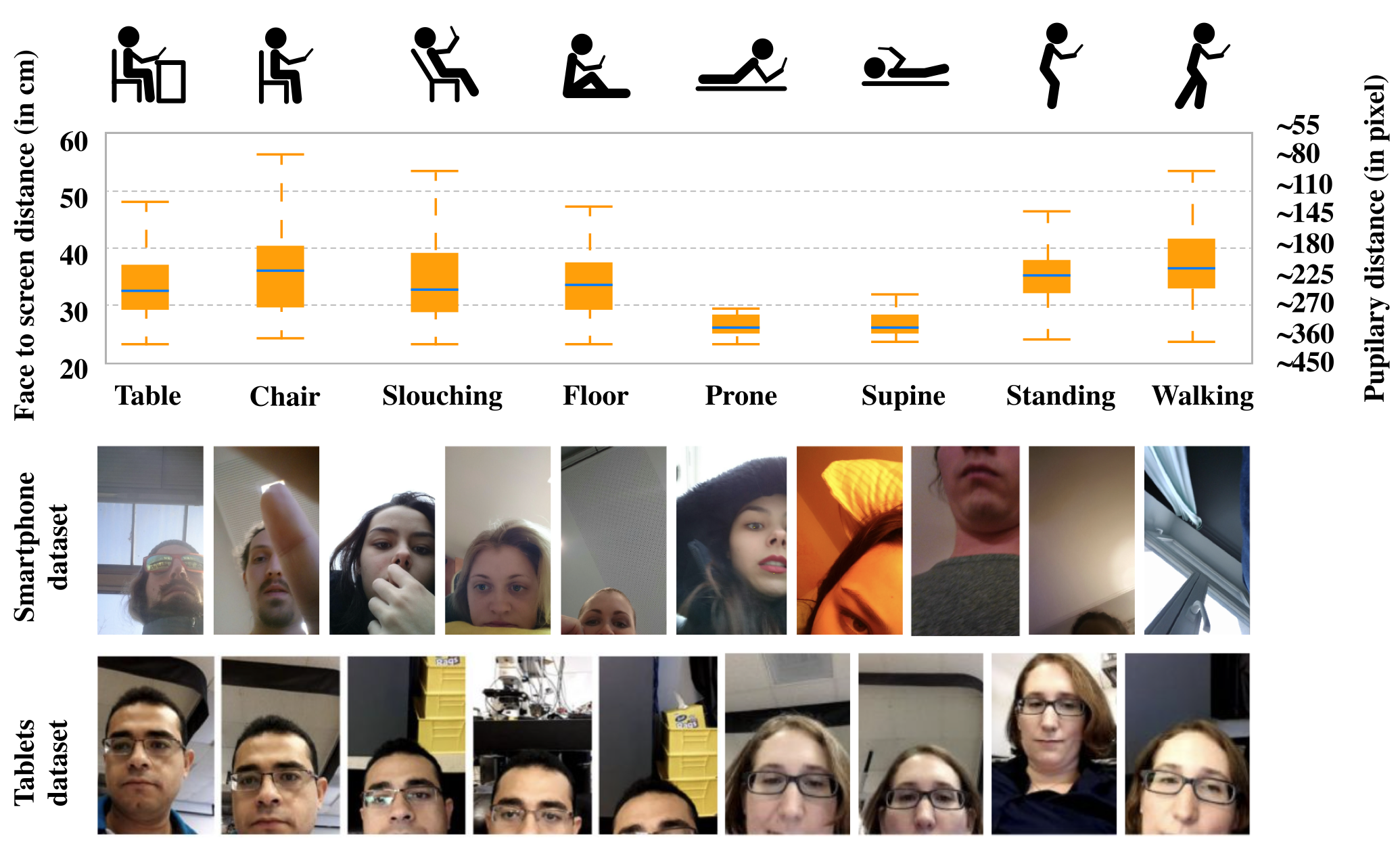}
    \caption{Various holding postures of handheld mobile devices and their impact on the distance between the face/pupils and the screen, and challenging examples for gaze estimation, images adapted from~\cite{huang2017screenglint, huang2017tabletgaze, khamis2018understanding, khamis2018past}}
    \label{fig:holdingposture}
\end{figure}

Commercial eye-tracking instruments, which generally come with data processing software such as Tobii Studio, process and transform raw eye-movement data into basic gaze events such as fixation, saccade, and saccade trajectory, and form scan path representation. These events are often defined by velocity, acceleration, and amplitude of gaze points and can be inferred via the following algorithms: velocity-threshold fixation identification (I-VT)~\cite{karn2000saccade, gitelman2002ilab, salvucci2000identifying} and identification by dispersion (I-DT)~\cite{kliegl1981reduction}, or spatio-temporal dispersion, the identification by dispersion and duration thresholds (I-DDT)~\cite{salvucci2000identifying, manor2003defining, krassanakis2014eyemmv}. 

I-VT aims to separate fixation and saccade from raw gaze points by calculating point-to-point velocity and acceleration, while I-DT and I-DDT achieve so by spatial dispersion; for example, fixation is identified when the speed of gaze points is low, and these points form in a dense cluster. The task of distinguishing between saccade, stationary and smooth pursuit is called a ternary classification task. Threshold-based algorithms for this task require a combination of several basic algorithms, such as Velocity Velocity Threshold Identiﬁcation (I-VVT)~\cite{komogortsev2013automated}, Velocity Dispersion Threshold Identiﬁcation (I-VDT)~\cite{komogortsev2013automated} and Velocity Movement Pattern Identiﬁcation (I-VMP)~\cite{komogortsev2010standardization}. These algorithms have worked well to distinguish saccade and fixation for commercial eye trackers in the lab setting, but due to their rigid threshold setting, they can result in a large error in detecting smooth pursuit~\cite{zhu2020hierarchical} when there is a variety in pupil sizes and viewing directions. The above algorithms supported in the open-source  libraries including PyGaze~\cite{dalmaijer2014pygaze}, EyetrackingR~\cite{dink2015eyetrackingr}, PyTrack~\cite{ghose2020pytrack}, Pupil~\cite{kassner2014pupil}, GazeParser~\cite{sogo2013gazeparser}, and GazeR~\cite{geller2020gazer}. Currently, a wide range of machine learning-based approaches is proposed to tackle this problem, including Bayesian~\cite{santini2016bayesian}, Random Forest~\cite{zemblys2018using}, Hidden Markov Models~\cite{zhu2020hierarchical}, DNN~\cite{zemblys2019gazenet}, CNN and LSTM~\cite{startsev20191d}.

However, for appearance-based gaze estimation on handheld devices, the applications only get the raw gaze points or directions, and there is no off-the-shelf software to process them into high-level patterns. The above algorithms might not be immediately applicable, either. The distance and angle between the head and handheld devices can constantly change, which can compromise the performance of these algorithms. Also, for many desktop environments, the user's gaze point is always within the screen, while for handheld devices, the cameras may only capture partial or occluded faces, or no faces at all (see Figure\ref{fig:holdingposture} adapted from \cite{huang2017screenglint, khamis2018understanding, huang2017tabletgaze,khamis2018past}). Therefore, an interesting research direction is to design and develop robust gaze pattern detection algorithms, resistant to fluctuated and erroneous gaze estimations. The analysis and processing of eye movement data need to combine segmentation and organisation at different scales depending on the purpose of the task; for example, from the original gaze path to the gaze event segmentation, and then to higher dimensional segmentation and organisation incorporating physiological and cognitive factors.

\subsection{Gaze Data Processing Pipeline}
Table~\ref{tab:eyedata} presents a general pipeline of processing gaze points from the area of eye tracking community~\cite{hein2017topology, duchowski2017eye}. It starts with data cleansing to improve the data quality, including noise reduction, de-nulling and dealing with outliers. Here, \textit{null} refers to missing data; for example, an eye-tracking device does not report position coordinates or a participant is not looking at the screen. This is often a necessary step for offline eye-tracking data analysis in desktop environments~\cite{hessels2015consequences, wass2014robustness}. Gaze event segmentation refers to detecting gaze events such as fixation or saccade from continuous gaze points using the algorithms in Section~\ref{subsec:eyedata}.

The intermediate data processing step is to visualise eye-movement data and remove the data that are beyond a reasonable range and merge data; for example, combining immediately adjacent fixation points. Further segmentation could be used to identify the area of interest (AOI)~\cite{hessels2016area}, inferring scan trajectories and heatmaps by saccade path and dwell time, as presented in Figure~\ref{fig:eyemovement}. Such information can serve as advanced features for applications.

\begin{table}
\centering
\caption{Eye movement data analysis}
\label{tab:eyedata}
\begin{tabular}{l|l} 
\hline
\multicolumn{1}{c|}{Pipeline} & \multicolumn{1}{c}{Description} \\ 
\hline
\hline
Data Clean & Noise Reduction, De-Nulling,  Eliminate Outlier and Blink, Smoothing \\
\hline
\begin{tabular}[c]{@{}l@{}}Segmentation for \\ Basic Gaze Events\end{tabular} & \begin{tabular}[c]{@{}l@{}}Fixation, Saccade, Smooth Persuit;\\ Fixation Statistics, Saccade Velocity, Event Duration\end{tabular} \\
\hline
Data Processing & \begin{tabular}[c]{@{}l@{}}Removal of Implausible Movements, Merging Intra-Threshold Movements \\Data Visualisation\end{tabular} \\
\hline
\begin{tabular}[c]{@{}l@{}}Segmentation for \\Further Purposes\end{tabular} & \begin{tabular}[c]{@{}l@{}}Area of Interest (AOI), Scan Path Representationl;\\Gaze Pattern Modelling: Dwell, Pursuit, Gaze Gesture, etc\end{tabular} \\
\hline
Possible Inferences & \begin{tabular}[c]{@{}l@{}}Modelling: Top-Down or Bottom-Up, Stochastic Processes; \\ Inferences: Personal Attributes, Cognitive Process, Attention \& Intention\end{tabular} \\
\hline
Possible Applications & UI Control \& Adaption, Medicine \& Healthcare, User Security \& Privacy, etc. \\
\hline
\end{tabular}
\end{table}

\subsection{Summary}
In the process of investigating gaze analytics and the development of gaze-based applications for handheld devices, we have uncovered several key findings that can guide future research and application design in this area.

Understanding the nuances of eye physiology and human visual attention mechanisms is crucial for the effective design of gaze interactive applications. By comprehending how different types of eye movements, such as saccades, fixations, and smooth pursuits, are related to human consciousness levels and attention mechanisms (bottom-up and top-down), researchers and designers can develop more ergonomic and intuitive gaze-based interaction. This understanding can also facilitate predictive reasoning about users' intentions and interests, which has a significant potential for creating more personalised and effective gaze interaction.

The development of robust gaze pattern detection algorithms is a critical research direction. Current algorithms are faced with the challenges such as changing distance and angle between the head and the device, partial or occluded face captures, and fluctuating gaze estimations. These factors can cause instabilities in eye movement data, leading to issues such as reduced frequency, incoherent absence, and other inconsistencies in gaze pattern detection. Therefore, creating algorithms that can withstand these variations and provide accurate gaze event segmentation and organisation is essential for the wider adoption of gaze-based applications on handheld devices.

\section{Gaze Interaction}\label{sec:application}
\begin{figure}[!htbp]
    \centering
    \includegraphics[width=0.95\textwidth]{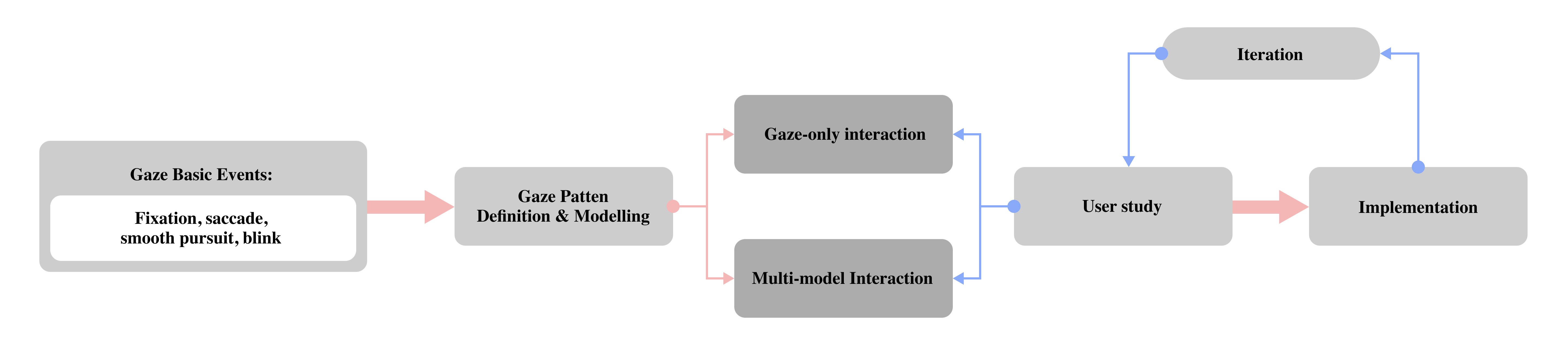}
    \caption{The workflow of gaze interaction}
    \label{fig:GazeFlow}
\end{figure}

This section will review the existing gaze-based interactive applications. We first introduce the categorisation of gaze interactions and then describe applications across different platforms, including handheld mobile devices, desktops, Virtual Reality (VR) and Augmented Reality (AR). These applications help to uncover new opportunities for interactive applications on handheld mobile devices. Figure~\ref{fig:GazeFlow} illustrates the overall design flow of a gaze-engaged interaction from eye movement to final testing and optimisation.

\subsection{Gaze Interaction}
Gaze-only interaction can be broadly classified into two main types: implicit and explicit. Implicit interactions involve the interface adapting to the user's passive gaze behaviour, while explicit interactions require the user to intentionally move their eyes to provide direct input. Implicit interaction is usually tailored to specific tasks, devices, and environmental characteristics. For instance, a reading application might predict users' reading speed and automatically turn pages by implicitly analysing users' gaze trajectories~\cite{kumar2007gazeenhance, lei2023DynamicRead}, or a display system could alter content rendering based on users' intentions~\cite{wilson2018autopager}. Such predictions are often achieved by applying machine learning techniques to estimate gaze behaviours and infer human intentions.

Explicit interaction, on the other hand, relies on users' voluntary and intentional gaze movements for manipulation. This type of interaction typically employs dwell time, pursuit, gaze gestures, or a combination of these techniques, which will be detailed below.

\subsubsection{Dwell-time}
Dwell time involves a brief fixating on a target for a period of time~\cite{paivi2009fast, dybdal2012dwell} in order to differentiate between casual viewing and gaze input. This addresses the Midas touch problem~\cite{jacob1991use} where users unintentionally gaze over the potential target and make selections. This technique has been widely used in gaze-only interactions and is useful for interaction techniques that do not require precise gaze estimates, e.g., gaze typing~\cite{paivi2009fast, Mott2017improvingDwell}. The recommended threshold for the dwell time is between 200 and 1500 ms for remote eye-tracking system~\cite{paivi2009fast, Mott2017improvingDwell}, and the exact threshold often requires trials-and-errors on the threshold for a specific task.

\subsubsection{Pursuit}
Pursuit refers to smooth pursuit eye movements where the eyes follow a moving object~\cite{robinson1986model, vidal2013pursuits, kowler2019predictive}. Pursuit measures the match between users' eye movement and the object's movement via Pearson correlation coefficient~\cite{Drewes2019DialPlates, vidal2013pursuits, Esteves2015Orbits, Velloso2017MotionCorr}, or machine learning techniques like CNN~\cite{startsev20191d} and Bayesian~\cite{santini2016bayesian}. 
It is often used as a calibration technique~\cite{zhe2020calibrationfreepursuit,Celebi2014pursuitCalib} as mentioned in Section~\ref{subsec:calibration}.

\subsubsection{Gaze Gesture}
Gaze gesture is a sequence of predefined eye movements (or called strokes)~\cite{Drewes2007InteractGesture, rozado2015controlling, heikkila2012simple, rajanna2018gaze}.  It is a promising alternative to other gaze interaction techniques, especially when the screen is too small to support other techniques~\cite{bace2016ubigaze}. One advantage of gaze gesture is that it can support a large number of commands using a small number of gesture combinations~\cite{Drewes2007InteractGesture}. However, the use of gaze gestures can introduce complexity, as users may have difficulty recalling complex gestures and initiating them physically~\cite{Majaranta2019GesturesIll}. There are machine learning-based methods such as Hierarchical Temporal Memory (HTM)~\cite{ROZADO2012Low, Rozado2012GestureRecognition} and Graph Neural Networks (GNN)~\cite{Shi2021GNNGesture} to detect and separate gaze gestures from noisy eye movement signals.

\subsection{Types of Gaze Interactions}\label{subsec:types}

Gaze-based applications are categorised into three groups based on the interaction and how gaze is acquired: explicit gaze interaction, implicit gaze interaction and multi-model gaze interaction~\cite{khamis2018past, majaranta2014eye}. The two main types of explicit gaze interaction applications are gaze typing and gaze interface control. The former allows users to input text via dwell time or gaze gesture~\cite{zhang2018efficient, Mott2017improvingDwell, paivi2009fast, rajanna2018gaze}, while the latter utilises people's voluntary eye movements and conscious gaze direction to control or communicate with a computer; for example, a user may perform simple horizontal or vertical eye movements to indicate disagreement or agreement~\cite{majaranta2014eye}. 

In implicit gaze interaction, there are three main application scenarios: attentive user interfaces, passive eye monitoring, and gaze-based user modelling. The attentive user interface applies users' natural eye movements rather than expecting particular gaze behaviours for explicit commands; for example, changing the movie plot based on the viewer's visual interest~\cite{tore2005gaze}. Gaze-based user modelling monitors and analyses the dynamics of gaze behaviour over time to understand users' behaviour, intention, and cognitive processes~\cite{zhang2019evaluation}. For example, eye movement patterns have been used to recognise human activities such as reading or common office activities~\cite{bulling2012multimodal}. The implicit gaze interaction signals can be used to recognise and speculate humans' latent behaviours, including measurement of users' preferences \cite{lagun2014towards}, attention \cite{newman2020turkeyes,faber2018automated,faber2020eye,mills2021eye}, interests \cite{li2017towards}, individual stress~\cite{huang2016stressclick}, emotional states~\cite{muller2018detecting}, and mental disorders such as schizophrenia and autism spectrum disorder~\cite{shishido2019application,babu2019understanding,elbattah2020nlp}. For example, Deng et al.~\cite{deng2020drivers} have analysed eye movement of drivers to predict their fixation and understand their attention allocation on scenes, and Pan et al.~\cite{pan2022lanchange} have explored the use of eye-tracking technologies to predict drivers' lane-changing intention. Passive eye monitoring is often for diagnostic applications~\cite{sink2020novel, duchowski2018gaze} where people's visual behaviours are recorded for offline processing~\cite{duchowski2018gaze}. 

Multi-modal gaze interaction refers to the use of gaze alongside other input modalities, such as voice, touch, or hand gestures. Multi-modal gaze interaction can overcome the limitations of other modalities when they are not accurate or when users have difficulty interacting with them. Gaze can also enhance other modalities; for example, gaze has been combined with touch-based PINs for a more secure authentication solution~\cite{khamis2017gazetouchpin}, and gaze is used to improve voice interaction by fixating on an object~\cite{mayer2020enhancing}. Both explicit and implicit gaze interactions can be employed in multi-modal interactions, offering a flexible and intuitive user experience.

In the context of handheld mobile devices, the applications mainly fall into the first two categories, and the trend is moving from gaze-only interaction to multi-model interaction. As listed in Table~\ref{tab:applications}, earlier applications have used explicit gaze gestures to perform particular commands. With the improvement of gaze detection, gaze has served as implicit interaction input; for example, deriving users' interest or attention by monitoring their gaze in the background~\cite{newman2020turkeyes, kar2020gestatten, ohme2021mobile}. Wrist-worn gaze control~\cite{hansen2016wrist} provides a gaze-based smart home control setting, which uses off-screen gaze gestures on a wrist-worn unit for IoT control. Gaze+Hold~\cite{gomez2021gaze+} provides a gaze-based interface, where a user can use gaze gesture combined with blink and fixation to complete most of the mouse functionality. For example, the selection of an object is performed by first fixating on an object and then closing and opening one eye. The downside of this type of application is that the gaze input might be slow and less accurate than the single-source input, and users may potentially experience fatigue~\cite{gomez2021gaze+}.

Also, the applications can be classified as gaze-only and multi-modal interaction~\cite{khamis2018past}. The former uses gaze solely as the interaction input, and this is widely adopted by gaze applications in handheld mobile devices. The latter uses gaze to complement other interaction modalities, such as touching or tilting. Pfeuffer and Gellersen \cite{pfeuffer2016gaze} have explored the gaze and touch interaction to extend the area that the thumb cannot reach. For example, a user can look at an element on the screen and tap anywhere, and the system will activate the element. GTmoPass~\cite{khamis2017gtmopass} is an authentication method that uses handheld mobile devices to enter multi-modal passwords; that is,  combining gaze gestures (e.g., left to right) with touch input. 

\begin{figure*}
\centering
\includegraphics[width=0.9\textwidth]{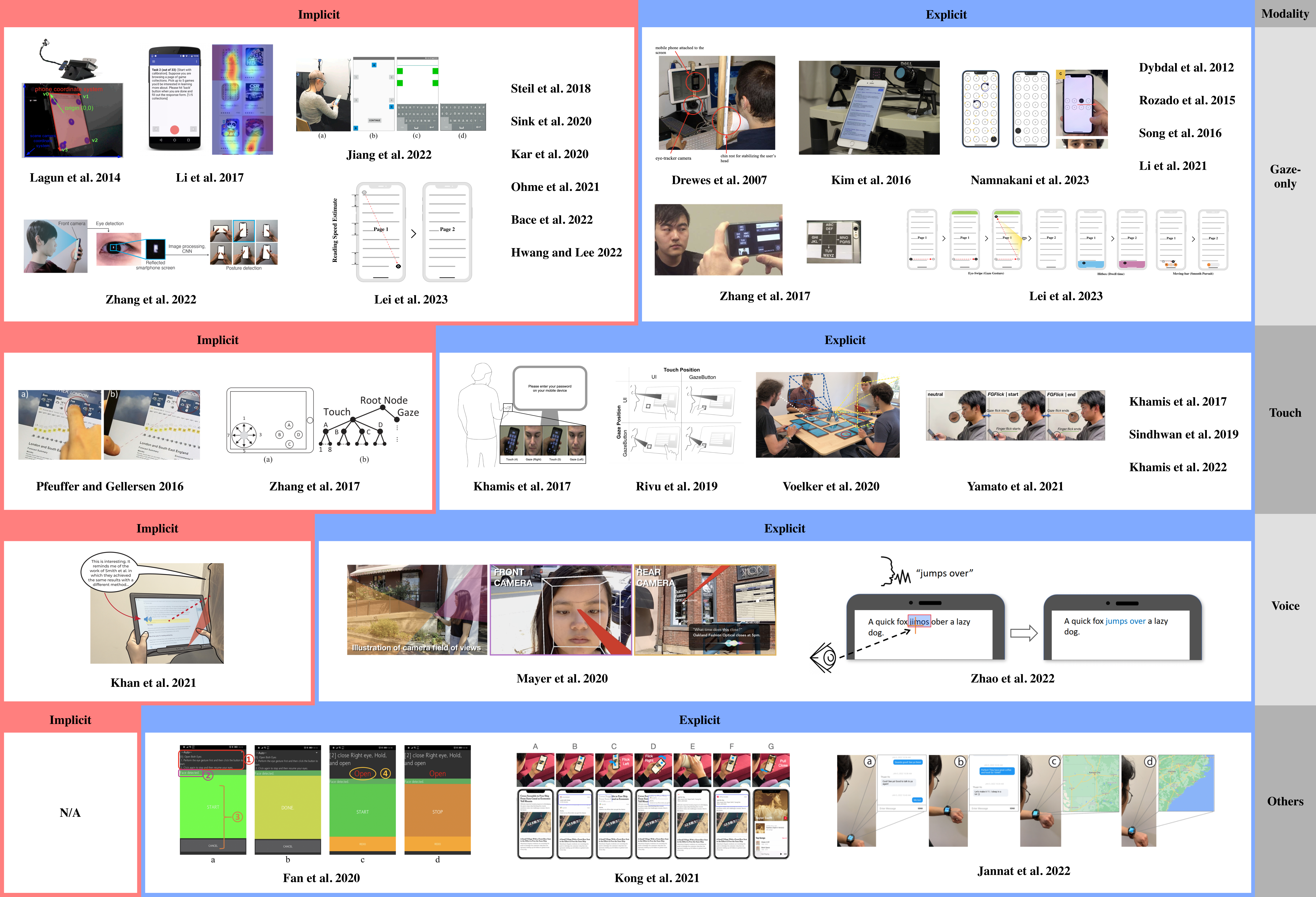}
\caption{Examples of gaze applications on handheld mobile devices}
\label{fig:applications}
\end{figure*}

\begin{table}
\centering
\caption{Gaze applications on handheld mobile devices}
\label{tab:applications}
\begin{tabular}{l|llcccl} 
\hline
\multicolumn{1}{c|}{Project} & \multicolumn{1}{c}{Year} & \multicolumn{1}{c}{Sensor} & Feature & Modality & Leverage & \multicolumn{1}{c}{Application} \\ 
\hline
\hline
         \cite{drewes2007eye} & 2007 & Eye Response ERICA  &gaze gesture &- &E & interface control \\
         \cite{dybdal2012dwell} & 2012 &external IR cam &dwell \& gesture &- &E & target selection \\
         \cite{lagun2014towards} & 2014 &Tobii X60 &dwell &- &I & attention analysis \\
         \cite{rozado2015controlling} & 2015 & modified prototype &gaze gesture &- &E & interface control \\
         \cite{pfeuffer2016gaze} & 2016 &Tobii EyeX &dwell &touch &I & interface control \\
         \cite{song2016eyeveri}& 2016 &phone camera & gaze basic events &- &E & user authentication\\
         \cite{kim2016pagination}& 2016 & Facelab 5 &gaze basic events &- &I & Website usability test  \\
         \cite{li2017towards}& 2017 & phone camera  &dwell &- &I & intention inference \\
         \cite{khamis2017gtmopass}& 2017 & phone camera &gaze gesture &touch &E & user authentication \\
         \cite{zhang2017exploring}& 2017 & Tobii eyeX &gaze basic events &touch & I & gaze adaptive UI \\
         \cite{zhang2017smartphone}& 2017 & phone camera & gaze gesture & - & E & gaze input\\
         \cite{khamis2017gazetouchpin}& 2017 & external RGB cam & gaze gesture &touch &E & user authentication\\
         \cite{steil2018forecasting}& 2018 & phone camera & gaze basic events &- &I & attention inference\\
         \cite{sindhwani2019retype}& 2019 & Tobii 4C & dwell  & touch & E & text editing aids\\ 
         \cite{rivu2019gazebutton} & 2019 & Tobii 4C & dwell & touch & E & text interface control\\ 
         \cite{mayer2020enhancing}& 2020 &phone camera & dwell  & voice & E & map navigation\\ 
         \cite{sink2020novel}& 2020 & phone camera & eye image &- &I & ocular exam\\
         \cite{voelker2020gazeconduits}& 2020 & phone camera & dwell & touch &E & cross-device control\\ 
         \cite{fan2020eyelid}& 2020 & phone camera & dwell  & eyelid & E &  interface control\\ 
         \cite{kar2020gestatten} & 2020 & phone camera & gaze basic events  & - & I &  attention analysis \\
         \cite{yamato2021fgflick}& 2021 & phone camera & gaze gesture & touch & E & gaze-assist input \\
         \cite{kong2021eyemu}& 2021 & phone camera & dwell & hand motion & E & interface control\\
         \cite{ohme2021mobile}& 2021 & Tobii X2 & gaze basic events & - & I &  attention analysis\\ 
         \cite{li2021bayesgaze}& 2021 & phone camera & dwell & - & E & target selection\\ 
         \cite{khan2021gavin}& 2021 & Tobii 4C & dwell & voice & I & implicit note-taking\\
         \cite{khamis2022user} & 2022 & external RGB cam & gaze gesture &touch &E & user authentication\\
        \cite{zhao2022eyesaycorrect}& 2022 & phone camera & dwell &voice &E & text correction\\
        \cite{bace2022privacyscout} & 2022 & external RGB cam & gaze \& face & - & I & user privacy \\
        \cite{zhang2022reflectouch} & 2022 & phone camera & eye image  & - & I & holding posture detection\\
        \cite{zhao2022don}  & 2022 & external RGB cam & gaze basic events  & * & E & gaze command definition\\
        \cite{hwang2022eye}  & 2022 & Tobii X2 & gaze basic events  & - & I & attention analysis\\
        \cite{jiang2022learning} & 2022 & SMI Glasses & gaze basic events  & - & I & learning process of typing\\
        \cite{jannat2022face}& 2022 & watch camera & face position & hand motion & E & spatial user interfaces\\
        \cite{namnakani2023comparing}& 2023 & phone camera & dwell, pursuit, gesture  & - & E & gaze UI usability test\\
        \cite{lei2023DynamicRead}& 2023 & phone camera & dwell, pursuit, gesture  & - & I\&E & gaze UI usability test\\
\hline
\end{tabular}
\begin{tablenotes}
      \small
      \item Note: \textit{Feature} means gaze feature for interaction; \textit{Modality} means the other modality with gaze; \textit{Leverage} means the way of leveraging gaze; \textit{E} means Explicit; \textit{I} means Implicit; gaze basic events mean: fixation, saccade, smooth pursuit etc; *: including eyelids, mouth, and head.
    \end{tablenotes}
\end{table}

\subsection{Wider Application Domains}\label{subsec:application_domains}
A large number of applications have employed gaze as an interactive modality to complement other modalities in different application domains, including accessibility and productivity, collaboration across devices, device and robot control, and security and privacy.

\subsubsection{Accessibility and Productivity}
Gaze can be an intuitive interaction modality to be complemented with other modalities to enhance accessibility and productivity. This area of applications can be further grouped into gaming~\cite{menges2017schau}, web browsing~\cite{kumar2017chromium}, typing~\cite{sindhwani2019retype}, and computer interaction~\cite{schenk2017gazeeverywhere,elmadjian2021gazebar,lewien2021gazehelp,rivu2019gazebutton}. The main idea is to detect users' attention and intention from their gaze, and use their gaze dwell time and gesture to trigger different actions. ReType~\cite{sindhwani2019retype}  is a gaze-assisted positioning technique, which makes use of users' gaze to position the text where they are interested. This interaction method reduces the use of the mouse and allows users to perform editing operations while keeping their hands on the keyboard. GazeHelp~\cite{lewien2021gazehelp} assists graphic design activities using real-time gaze information. It works as a plugin to Adobe PhotoShop to allow users to select tools with gaze, create windows at the gaze point, and block the current artboard when the gaze is away from the display. It integrates gaze points and a mechanical switch for object selection, positioning, and manipulation. 
Gaze has been applied to empower disadvantaged people; for example, creating visual design~\cite{creed2020multimodal} and coding~\cite{paudyal2020voiceye}.

\subsubsection{Collaboration}
Prior works also point out that gaze interaction can improve multi-user collaboration efficiency and user experience~\cite{pfeuffer2021multi,schlosser2018beyond, kutt2020effects, he2021gazechat}. Gaze-sharing user interface~\cite{schlosser2018beyond} detects the user's attention by sensing the user's eye movements and shares them with other collaborators to enhance team collaboration. GazeChat~\cite{he2021gazechat} is a remote communication system that renders gaze-aware 3D profile photos. It uses an off-the-shelf webcam to track users' gaze in video calls such as Zoom or Teams and then renders gaze to animate the participants' profile images. 

In addition to team collaboration, gaze can also support multi-device interaction. It is increasingly common for a user to have multiple devices or multiple screens. In a collaborative environment with one or more tablets, GazeConduits~\cite{voelker2020gazeconduits} uses the phone's front camera to sense and detect the user's gaze to identify which tablet the user is looking at and making content selections and actions based on the user's gaze gestures. GazeMirror~\cite{voelker2020gazeconduits} allows users to mirror the content between devices by coordinating gaze and four-finger multi-touch gesture. Another gaze+touch project~\cite{pfeuffer2021multi} allows multiple users to zoom in and out on a single shared map without interference or occlusion issues.

\subsubsection{Control and Interaction}
Gaze is playing an important role in controlling devices~\cite{jungwirth2018contour} and robots~\cite{li2020gaze}. Krishna et al.~\cite{krishna2020eyerobotic} propose a control interface that uses a tablet to process and estimate gaze. The project uses dwell time as a trigger for interaction instruction and manipulates the robot arm to perform a pick-and-drop operation. Jungwirth et al.~\cite{jungwirth2018contour} have designed gaze-triggered actions that use object contours as visual guidance; for example, a user can trace the contour of a lamp to turn on the light.

\subsubsection{Security and Privacy}
Gaze supports security and privacy in applications. GazeConduits~\cite{voelker2020gazeconduits} provides user awareness in a poker game; for example, showing playing cards when the gaze looks and hiding them when the gaze leaves. It also maintains the user's space around the table by detecting who enters and leaves the space at what time and accordingly adjusting the content display. GazeRoomLock~\cite{george2020gazeroomlock} combines gaze and head pose for user authentication in VR applications. EyeVeri~\cite{song2016eyeveri} applies signal processing and pattern matching techniques to explore conscious and unconscious gaze patterns for access authentication. In addition, gaze-based interaction can support authentication and privacy protection on personal devices~\cite{katsini2020role} such as password entry~\cite{khamis2017gtmopass} and protection against shoulder surfer~\cite{ragozin2019private}.

\subsubsection{Healthcare and Other Areas}
As an important bio-signal, eye movement patterns have been widely used in the diagnosis of mental health such as depression\cite{alghowinem2013eye}, Parkinson's\cite{harezlak2018application}, autism~\cite{venuprasad2019characterizing}, and dyslexia~\cite{rello2015detecting}. For example, autism can be diagnosed by analysing joint attention from interactions between eye movement and objects in a room~\cite{venuprasad2019characterizing}. Gaze is also utilised for assessing the usability of tools or systems~\cite{bace2019accurate,paulus2021usability}. Bace et al.~\cite{bace2019accurate} have employed an unsupervised approach to detect gaze contact and attempted to compute and quantify attentional metrics. This allows for analysing how users allocate attention during interactions.

\subsection{Summary}
We have reviewed a broader scope of applications that make use of gaze as an interaction modality. Most of these applications use commercial eye-tracking devices, including Tobii~\cite{elmadjian2021gazebar,lewien2021gazehelp} and Eye tribe~\cite{creed2020multimodal}.  We have seen an increasing number of applications for collaboration, control and interaction, security and authentication developed on mobile devices. In terms of accessibility and productivity, there is a tendency to explore complex actions from gaze, which can be challenging for handheld devices. For example, typing on the phone's keyboard often requires the gaze to switch back and forth between the target area and the keyboard. This makes it challenging to locate the gaze position accurately and thus can trigger the Midas touch problem. The key to tackling the Midas touch problem is to analyse the eye movement metrics through the cognitive process and separate users' true intention from the unintentional activities~\cite{parisay2020eyetap, elmadjian2021gazebar, festor2022midas, gomez2021gaze+}. The other applications that combine fixation and eye-opening/closing actions to select and manipulate objects can be further explored. A potential limitation is that the mobile devices have smaller icons that are presented closely together, which can make the estimation of gaze and inference of gaze trajectories more challenging.

\section{Future Research Challenges and Opportunities}\label{sec:future}
This section focuses on addressing key questions related to the challenges of handheld mobile devices in advancing unconstrained gaze estimation, robust gaze data processing methods and facilitating interaction involving gaze. Our discussion will be centred around three main questions: 
\begin{itemize}
    \item \textbf{RQ1}: How can we achieve robust gaze estimation in unconstrained environments?
    \item \textbf{RQ2}: How can we develop gaze analysis and processing methods that can tolerate the inherent instabilities of dynamic gaze estimation? 
    \item \textbf{RQ3:} How can we utilise estimated gaze for a broader range of applications?
\end{itemize}

\subsection{Roadmap for Unconstrained Gaze Estimation}
The mobile setting is bringing new challenges to gaze estimation. First of all, the distance and angle between the screen and the eyes might not be ideal or stable. The consequence is that the camera might only capture a partial face or an occluded face; for example, the face could be covered by sunglasses or a scarf (see Figure \ref{fig:holdingposture}). Secondly, users tend to change their holding postures, motion states, and whereabouts, which may compromise the gaze estimation and/or calibration model. Thirdly, the variety of environmental conditions, such as lighting may compromise the quality of images for face and eye detection. In the following, we will present potential directions to tackle these challenges towards achieving unconstrained gaze estimation. 

\subsubsection{Augmenting with Sensors and Multi-Cameras}
One of the major differences between mobile devices and desktop/AR/VR devices is that modern mobile devices are augmented with a rich set of sensors. These sensors provide more information that can reveal the context of the user. For example, we can infer the holding posture from the gyroscope, physical activities from the accelerometer, and the lighting condition from light sensors. Also, we have witnessed an improvement in the camera and sensor technologies on mobile devices. For example, some mobile devices have between three and six cameras with higher resolutions. Combining data from these built-in sensors with the camera may improve the accuracy of gaze estimation~\cite{chang2021high}.

\subsubsection{Continuous Calibration}
Calibration is often necessary to tune the gaze estimation model and to allow for more accurate estimation adapting to the current user, screen, and environment. \textit{Continuous calibration} is an interesting direction to explore; for example,  quickly adapting the model continuously and obtaining the calibration points without distracting users from their tasks at hand.

A potential direction is to allow for implicit calibration, which leverages users' interaction and other sensor data in mobile devices, including accelerometer, gyroscope, magnetometer, and compass. Gaze behaviour, such as smooth pursuit, has been explored for calibration. For example, correlating eye or hand movements with the trajectory of moving objects has been utilised for in-use calibration ~\cite{negulescu2012recognition,carter2016pathsync}. The trajectory of a straight saccade has been used to calibrate the distortion of eye tracker~\cite{huang2019saccalib}. 

Current handheld mobile devices integrate a number of inertial sensors, and their data can be employed for dynamic adjustment. Pino et al.~\cite{pino2012improving} propose to detect changes in the position and holding postures of the phone from gyroscopes and accelerometers. The detected change will inform whether to use current eye images for gaze estimation. Gaze-based input requires a high level of stability in the input process, and device movements and other possible disturbances can decrease the quality of input. A promising direction is to use inertial sensor data to select the input for gaze estimation algorithms or select and tune the gaze estimation algorithms for the current input.

Research can investigate real-time feedback and calibration mechanisms to enhance the resilience of gaze estimation in handheld mobile devices. Real-time feedback on users' gaze behaviour can be used to calibrate the model on-the-fly, leading to high resilience even in the face of changing conditions.

\subsubsection{Diversity in Datasets}
As presented in Section~\ref{sec:datasets}, current datasets predominantly contain static poses and are limited in terms of capturing gaze estimation in dynamic conditions. There is a need to collect a large amount of data with versatile conditions in more naturalistic settings. However, such data can be expensive to collect; therefore, the current research direction is to explore unsupervised~\cite{guo2020domain}, self-supervised~\cite{cheng2018appearance, wu2022gaze} and weakly supervised techniques~\cite{kothari2021weakly}.

One increasing concern is \textit{bias} in face recognition models; for example, the state-of-the-art models are trained on the datasets that over-represent socio-demographic groups with certain skin tones and facial structures, which makes them less accurate for marginalised groups~\cite{Wehrli2021}. For example, Buolamwini and Gebru have evaluated three commercial gender classification systems and found that the error rate for classifying lighter-skinned males is 0.8\% while the error rate for classifying darker-skinned females can reach up to 34.7\%~\cite{buolamwini2018gender}. The bias in the dataset has led to low-quality facial feature extraction on under-represented groups, which can result in low accuracy in gaze estimation. Therefore, future data collection should cover a wider ethnic group of subjects.

\subsubsection{Model Deployment}
Deploying high-performance gaze estimation models on handheld mobile devices presents challenges in model compression and model deployment.

\textit{Model Compression.} The existing gaze estimation models, especially the ones based on deep learning, can be computationally expensive and take up much memory; for example, a deep learning model can have billions of parameters. Even though today's smartphones have much more computational resources, continuously estimating gaze in real-time can still be challenging, and it can compromise their battery life and reduce user experience. Model compression is a popular direction to reduce the size of deep learning models while maintaining model performance. There are several approaches, such as parameter quantisation, parameter pruning, low-rank factorisation and knowledge distillation. Guo et al.~\cite{guo2019generalized} apply knowledge distillation and pruning to reduce the size of the CNN model while maintaining its performance. 

\textit{Model Deployment.} Many of the existing gaze estimation models are deployed on the cloud or edge so that mobile devices only perform pre-processing steps and then pass the extracted features to the deep learning models. However, this adds to the communication cost, results in high latency, and increases privacy risk. Future work will look into deploying the models on users' own devices. However, this raises a practical issue -- model framework compatibility. 

Most of the deep learning models are implemented in frameworks such as PyTorch or TensorFlow, and these frameworks do not have the same support on different handheld devices. The ONNX (Open Neural Network Exchange)~\cite{bai2019onnx} can be used as a medium to transform models across different deep learning frameworks and thus can help deploy models to various platforms of handheld mobile devices. It only supports operators that are commonly supported by all the deep learning frameworks, but not for specialised operators for specific frameworks. This limits the deployment of advanced, customised, gaze estimation models on handheld devices. An engineering perspective of research is how to support gaze estimation on a wide range of different devices. 

\subsubsection{Open Standards}
It is important to highlight the current lack of international or national industrial standards specifically for eye-tracking technologies and systems. Existing standards such as \textit{ISO 9241-971:2020 Ergonomics of human system interaction} and \textit{ISO 13407:1999: Human-centred design processes for interactive systems} focus on usability assessment and human-centred design principles. Similarly, \textit{WCAG 2.1: Web Content Accessibility Guidelines} is for conducting usability studies with people with disabilities, which are relevant to the use of oculomotor systems.

There is a need for the industry to develop appropriate standards that regulate the overall performance and usability of eye-tracking systems. These standards should address various aspects, including device accuracy, calibration procedures, data processing, and user privacy. Establishing such standards could ensure consistent quality and interoperability across different eye-tracking systems, facilitating their adoption and fostering innovation in the field.

\subsection{Roadmap for Gaze Analytics}
As described in Section~\ref{sec:gazeanalysis}, there are no established algorithms and toolkits for processing raw gaze points or vectors into high-level gaze patterns or events. Appearance-based gaze estimation on handheld mobile devices can be much less accurate than commercial eye-tracking devices in terms of large error, low frequency, and high instability. The existing methods of removing outliers, smoothing data, or merging intra-threshold movements might lead to significant information loss, compromising the quality of gaze analytics. Therefore, a future research direction is to develop gaze event detection and processing algorithms to tolerate imperfect gaze estimation. This may require a further understanding of user behaviour patterns and cognitive processes in various applications. Overcoming these challenges will not only facilitate more robust and accurate gaze interaction but also pave the way for a wide range of novel applications in human-computer interaction.

\subsection{Roadmap for a Wider adoption of Gaze Applications}

\subsubsection{Privacy Implication}
Gaze interaction can have high privacy implications. First of all, gaze can reveal users' intentions, interests, emotions, and personality traits~\cite{barz2020visual, wang2021scanpath, hoppe2018eye,partala2000pupillary}. Secondly, gaze estimation from handheld mobile devices is based on users' facial images. Thirdly, the camera may capture the background, such as bystanders' activities. Research effort should be devoted to protecting the foreground and background users' privacy; for example, how to securely store face/eye images, prevent leaking gaze information to third-party applications on the same devices, and explore the possibility of protecting the privacy of bystanders~\cite{katsini2020role}.

\subsubsection{Gaze Interaction}
One obstacle to the wider adoption of gaze interaction on handheld mobile devices is the imprecision of estimation~\cite{barz2018error} and sensitivity to the environments and interaction positions. Imprecision can increase users' fatigue and lower user experience. Furthermore, the screen size of handheld mobile devices might limit the choices of gaze pattern combinations; for example, saccade detection on a small screen can be challenging and smooth pursuit or dwell and gesture combination might be a better option.

Promising applications for gaze interaction on handheld mobile devices include healthcare, usability testing, cognitive process understanding, interface and device control, and security and privacy protection. For example, gaze-engaged interface or device control in accessibility contexts can offer an alternative input method for users with motor impairments, and gaze patterns can be used as bio-metric identifiers to enhance security and privacy protection.

Apart from using the front camera to capture the gaze behaviour of the target user, leveraging other cameras, such as rear or external cameras, to capture and project multiple gaze directions and identify objects of interest or perform interaction~\cite{kellnhofer2019gaze360} has rarely been explored. This approach holds promise for opening up a new research area, as it has the potential to link gaze with physical objects or locations in the real world through socialised gaze cues, enabling a more immersive user experience. Gaze-engaged multi-modal interaction is another promising area, as exploring such interactions can complement gaze and other modalities, leading to novel human-computer interaction.

\section{Conclusion}\label{sec:conclusion}
Mobile human-computer interaction is one of the most popular areas for innovation.  This paper has presented a review of gaze estimation and interaction on handheld mobile devices. There are many promising developments in this area: more powerful mobile devices with high-resolution cameras, accurate gaze estimation algorithms based on deep learning, and an increasing number of novel applications that leverage the use of gaze to enable hands-free interaction or complement other interaction modalities. This paper summarises these latest developments and points to the future research challenges and opportunities, especially in gaze estimation accuracy in terms of robustness and continuous calibration, the computational cost in terms of model size and power consumption, and gaze interaction in terms of richer eye movement exploration. It is desirable that the gaze estimation algorithm is adaptable to different environmental conditions and has a low computational overhead. Continuous and implicit calibration will be key factor in supporting unconstrained gaze estimation.

Gaze estimation algorithms are becoming increasingly mature and, although still far from unconstrained estimation, can be applied on handheld mobile devices by installing apps that deploy models.  We review the existing types of gaze interactions, summarise various types of eye movements, cognitive theories and combinations of gestures used in gaze interactions, and give a generalised workflow of gaze interactions in the hope that more people will see the role of gaze in interactions and use it to develop more and novel interactions and applications.

% ======== Main Content ========%

% \begin{acks}
% \end{acks}

\bibliographystyle{ACM-Reference-Format}
\bibliography{reference}

\appendix

\received{May 2022}
\received[revised]{April 2023}
\received[accepted]{June 2023}

\end{document}